\documentclass[journal]{IEEEtran}
\pdfoutput=1
\usepackage{amsmath,amsfonts,amssymb, amsthm}
\usepackage{array}
\usepackage{booktabs}
\usepackage{multirow}
\usepackage{textcomp}
\usepackage{stfloats}
\usepackage{url}
\usepackage{verbatim}
\usepackage{balance}  
\usepackage{graphicx}
\usepackage{cite}
\usepackage{subcaption}
\usepackage{algorithm}
\usepackage{adjustbox}
\usepackage{algpseudocode}
\usepackage{caption}
\usepackage[normalem]{ulem} 
\usepackage[table,xcdraw]{xcolor}
\usepackage[hidelinks, bookmarks=false]{hyperref}
\usepackage{xcolor} 

\pdfoutput=1
\begin{document}
\captionsetup[figure]{font={small}, name={Fig.}, labelsep=period}

\title{BeamCKM: A Framework of Channel Knowledge Map Construction for Multi-Antenna Systems}

\author{
	Haohan~Wang,
	Xu~Shi,~\IEEEmembership{Member,~IEEE},
	Hengyu~Zhang,
	Yashuai~Cao,~\IEEEmembership{Member,~IEEE},
	Sufang~Yang,
	Jintao~Wang,~\IEEEmembership{Senior~Member,~IEEE},
	Kaibin~Huang,~\IEEEmembership{Fellow,~IEEE}
	\thanks{
		Part of this work \cite{wang2025beamforming} will be presented at the IEEE Globecom 2025 in Taipei, 8-12 December 2025. \textit{(Corresponding author: Xu Shi)}
		
		Haohan Wang, Xu Shi, Hengyu Zhang, and Jintao Wang are with the Department of Electronic Engineering, Tsinghua University, Beijing 100084, China and Beijing National Research Center for Information Science and Technology (BNRist). (e-mail: \{whh24@mails., shi-x@mail., zhanghen23@mails. , wangjintao@\}tsinghua.edu.cn).
		
		Yashuai Cao is with the School of Intelligence Science and Technology, University of Science and Technology Beijing, Beijing 100083, China (e-mail:	caoys@ustb.edu.cn).
		
		Sufang Yang is with the Future Research Laboratory,
		China Mobile Research Institute, Beijing 100053, China (e-mail: yangsufang@chinamobile.com).

		Kaibin Huang is with the Department of Electrical and Electronic Engineering, the University of Hong Kong, Hong Kong (e-mail: huangkb@eee.hku.hk).

	}
}

\maketitle

\begin{abstract}
The channel knowledge map (CKM) enables efficient construction of high-fidelity mapping between spatial environments and channel parameters via electromagnetic information analysis. Nevertheless, existing studies 
are largely confined to single-antenna systems, failing to offer dedicated guidance for multi-antenna communication scenarios. 
To address the inherent conflict between traditional real-value pathloss map and multi-degree-of-freedom (DoF) coherent beamforming in B5G/6G systems, this paper proposes a novel concept of BeamCKM and CKMTransUNet architecture. The CKMTransUNet approach combines a UNet backbone for multi-scale feature extraction with a vision transformer (ViT) module to capture global dependencies among encoded linear vectors, utilizing a composite loss function to characterize the beam propagation characteristics. Furthermore, based on the CKMTransUNet backbone, this paper presents a methodology named $\text{M}^3$ChanNet. It leverages the multi-modal learning technique and cross-attention mechanisms to extract intrinsic side information from environmental profiles and real-time multi-beam observations, thereby further improving the map construction accuracy.
Simulation results demonstrate that the proposed method consistently outperforms state-of-the-art (SOTA) interpolation methods and deep learning (DL) approaches, delivering superior performance even when environmental contours are inaccurate. For reproducibility, the code is publicly accessible at https://github.com/github-whh/BeamCKM.
\end{abstract}


\section{Introduction}
\subsection{Background}
For the sixth-generation (6G) communication system, real-time and precise channel state information (CSI) is indispensable to realizing the paradigm shift toward ubiquitous, intelligent, and high-performance communication networks \cite{6GVision, Advancing6G, HigherSpectral}. However, the rapid advancements of array aperture have introduced huge difficulties in acquiring CSI due to the increasing pilot overhead, such as extremely large-scale multiple-input multiple-output (XL-MIMO) \cite{XL-MIMO, SpatialChirp}, reconfigurable intelligent surface (RIS) \cite{TripleStructured}, and other millimeter-wave (mmWave)  techniques \cite{mmWaveTHz}. Consequently, the dual pursuit of high-precision and low-latency CSI remains a prominent and unresolved challenge in multi-antenna communication scenarios.
To address this, channel knowledge map (CKM) has been proposed as an innovative environment-aware framework for CSI acquisition. As a high-fidelity portrayal of the propagation environment, CKM serves as a cornerstone for advancing future communication networks, supporting essential capabilities like spectrum sensing, beam management, and channel estimation \cite{CKM}. By the mapping between spatial location information and multi-dimensional channel parameters, CKM holds substantial promise for delivering key insights such as delay, angle of arrival (AoA), angle of departure (AoD), and pathloss, thereby effectively addressing the inherent limitations of traditional CSI acquisition schemes.

Despite its promising potential, the construction of precise CKM currently faces significant challenges, particularly in multi-antenna systems with multiple degrees of freedom (DoFs). \textit{Firstly}, existing studies are largely confined to single-antenna systems, failing to offer dedicated guidance for multi-antenna communication scenarios. \textit{Secondly}, an inherent contradiction arises from the use of incoherent one-dimensional real-valued maps for characterizing coherent multi-DoF beamforming systems. \textit{Thirdly}, current CKM construction heavily depends on environmental contours, which suffer from inaccuracies due to dynamic environments and GPS-introduced location errors, thereby compromising the reliability of the generated CKMs. Additionally, investigations into CKM construction utilizing observational data (i.e., real-time sensor measurements) are relatively underdeveloped, and the integration of multi-modal observations from diverse beams has been largely overlooked in existing literature. Hence, developing a robust methodology to construct accurate and reliable CKMs for multi-antenna systems is critical.

\subsection{Related Works}
CKM construction originated with the application of ray tracing \cite{Raytracing}, which is widely acknowledged as a feasible method for pathloss modeling with excessive computational complexity. To mitigate this, mathematics-based techniques, such as the Kriging algorithm \cite{Kriging}, dynamic mode decomposition \cite{DMD}, and matrix completion \cite{Matrix}, have been utilized to reduce the computational cost. However, these methods suffer from performance degradation as the number of sparse samples decreases. More recently, RadioUNet \cite{RadioUnet} pioneered the first neural network-based prediction approach that relies solely on BS positions and environmental contour maps, significantly simplifying the generation of radio map. With continuous advancements in computer vision and deep learning (DL), subsequent methods based on generative adversarial networks (GANs) \cite{GAN} and diffusion models \cite{WiFi,RadioDiff} have been proposed, further enhancing the accuracy of pathloss predictions. Beyond these mainstream architectures, the approach in \cite{I2I} utilized a Laplacian pyramid (LP) mechanism for constructing CKM, which incorporated multi-scale processing and significantly improved the inference speed. Meanwhile, the PMNet \cite{PMNet} with atrous convolution and hourglass networks has also demonstrated outstanding construction accuracy.
However, most existing methods are limited to single-antenna systems, which limit their applicability in practical multi-antenna communication systems. The fundamental shortcoming lies in their inability to capture and incorporate spatial channel characteristics, such as AoA and AoD, which are essential for effective beamforming. 
This limitation further restricts their applications in emerging technologies, such as RIS and integrated sensing and communication (ISAC) systems \cite{ISAC}.

Nowadays, research on multi-antenna CKMs remains limited and faces significant challenges. For instance, though channel angle map (CAM) and beam index map (BIM) were fully exploited in \cite{CAM} to reduce the prohibitive training overhead in MIMO systems, the corresponding CKM construction process was not thoroughly considered. CKMImageNet \cite{Dataset} provided a database including various MIMO parameters. However, it was based on an ineffective ray-tracing approach with excessive computational complexity. To address this issue, \cite{Mapping} derived a problem-specific neural architecture to map channel parameters and user positions, highlighting the explainability of the learning paradigm, yet its requirement for densely sampled data from a specific environment limited real-world scalability.
In \cite{Reduced}, the generation of CKM was implicitly integrated into the beam design module via location-dependent beamforming. While this method could effectively reduce the pilot overhead of beam management, its application was overly constrained, precluding the generalization and deployment of CKM to other communication modules.

From a theoretical perspective, the implementation of CKM construction for massive MIMO systems faces numerous challenges: 1) Phase ambiguity arises inherently from limitations in CKM resolution and unavoidable positioning errors, which deteriorates the coherent beamforming of massive MIMO; 2) A prominent issue in the CKM construction lies in the insufficient continuity and consistency during the formulation of its key parameters. Specifically, key parameters such as propagation delay, AoA/AoD, and pathloss exhibit fundamentally distinct physical characteristics (e.g., delay is time-domain dependent, angle is spatial-domain dependent, and pathloss is largely related to signal attenuation) and dynamic variation patterns; 3) Across geographical sampling points, dimension mismatch and storage disorder frequently occur among multiple propagation paths. To the best of our knowledge, no prior work has perfectly addressed the CKM construction of massive MIMO systems.

\subsection{Contribution}
This paper addresses two main obstacles in multi-antenna CKM construction. 
\textit{Firstly}, due to the limitations in map resolution and user location accuracy, it is impractical to construct maps that incorporate phase parameters. As a result, we can only utilize non-coherent parameters for map construction. However, beamforming inherently involves phased array coherence. To resolve this issue, we propose measuring pathloss under different beamforming vectors, thereby eliminating phase ambiguity and simplifying storage. 
\textit{Secondly}, considering the practical implementation of finite static beamforming codebooks in realistic MIMO architectures, we develop a codebook-specific methodology that demonstrates both practical applicability and operational effectiveness. This approach facilitates the reduction of equivalent channel representation to pathloss characterization across individual configurations. The principal contributions of this paper are detailed as follows:

\begin{itemize}
	\item We propose the concept of BeamCKM to address the critical DoF mismatch arising between real-value (non-coherent) CKM and coherent beamforming transmission, which reformulates the traditional one from $\mathcal{F}(\mathcal{P}|\boldsymbol{E})$ into an extended beam-dependent representation $\mathcal{F}(\mathcal{P}; \boldsymbol{f}|\boldsymbol{E})$, which offers benefits in terms of beam management and interference mitigation.
	\item We present one BeamCKM construction approach named CKMTransUNet, by leveraging its inherent radial propagation characteristics and spatial environmental properties. In detail, we integrate CNN-based feature extraction, UNet-driven hierarchical spatial reconstruction, and vision transformer (ViT)-enabled non-local interaction of environmental layouts, thereby forming a synergistic design to boost channel representation  performance. Besides, we develop a composite loss to further capture the characteristics of beam propagation, wherein the Sobel operator and LP structure are adopted for edge extraction and multi-scale spatial information, respectively.
	\item To overcome the distortion of outdated environmental profiles, we develop a multi-modal BeamCKM construction scheme named $\text{M}^3$ChanNet with a balance between fixed environmental contours and sparse real-time observations. To realize this, $\text{M}^3$ChanNet leverages the CKMTransUNet backbone to extract features from environment and employs a multi-modal cross-attention mechanism to merge all feasible information from multi-beam observations. 
	\item We provide the publicly available BeamCKM dataset comprising diverse urban structures. Using the Sionna ray-tracing simulator, we systematically place BSs in various scenarios under different codewords, thereby establishing a sufficiently large and widely covered high-quality BeamCKM training dataset. Meanwhile, experimental results demonstrate that the proposed multi-modal multi-beam CKM network consistently outperforms all state-of-the-art (SOTA) methods.
\end{itemize}

\subsection{Organization}
The remainder of this paper is organized as follows. Section II introduces the system model and CKM construction procedure. In Section III, we develop the BeamCKM framework for massive MIMO. In Section IV, we present  CKMTransUNet approach along with composite loss architecture. Section V details the methodology for BeamCKM construction from sparse observations. Simulation results are analyzed in Section VI. Finally, Section VII provides conclusions and future research directions.

\textit{Notation}: Lower-case and upper-case boldface letters denote column vectors and matrices, respectively. The conjugate transpose is denoted by $(\cdot)^{H}$. The $L_1$ norm and the Euclidean norm of a vector $\mathbf{a}$ are denoted by $\|\mathbf{a}\|_{1}$ and $\|\mathbf{a}\|_{2}$, respectively. The Hadamard product is signified by $\odot$. The indicator function $\mathbb{I}(\cdot)$ returns 1 if the condition is true and 0 otherwise.

\section{System Model}
This section first introduces the channel model for MIMO systems. Subsequently, we present the traditional single-antenna CKM construction framework.

\subsection{Channel Model}
We consider a downlink MISO system where the BS is equipped with $N_{BS}$ antennas arranged in a uniform linear array (ULA) and serves multiple single-antenna users. The transmitted symbol for user $i$ is expressed as $s_i$, so the data sent from BS can be written as $\mathbf{x}_i = \boldsymbol{f}_is_i$, where $\boldsymbol{f}_i \in \mathbb{C}^{N_{BS} \times 1}$ is the precoding vector for the $i$-th user with $||\boldsymbol{f}_i||_2 = 1$. Supposing the channel vector between the BS and the $i$-th user at position $\mathbf{p}_i = (x_i, y_i)$ is expressed as $\boldsymbol{h}_i \in \mathbb{C}^{N_{BS} \times 1}$, the received signal is given by:
\begin{equation}
	y_i = \boldsymbol{h}^H_i\boldsymbol{f}_i s_i + n_i,
\end{equation}
where $n_i \sim \mathcal{CN}(0, \sigma_N^2)$ denotes additive white Gaussian noise (AWGN). The channel $\boldsymbol{h}_i$ can be further expressed as:
\begin{equation}
	\boldsymbol{h}_i = \boldsymbol{h}_i^{\mathrm{LoS}} + \sum_{l=1}^{L} \boldsymbol{h}_{i,l}^{\mathrm{NLoS}},
\end{equation}
where $\boldsymbol{h}_i^{\mathrm{LoS}}$ denotes the \emph{line-of-sight} (LoS) component corresponding to direct propagation, $\boldsymbol{h}_{i,l}^{\mathrm{NLoS}}$ represents the $l$-th \emph{non-line-of-sight} (NLoS) component arising from electromagnetic wave reflections and scattering \cite{38901}. While this model provides detailed channel characterization, it is limited due to the large number of parameters. To address this limitation, we propose a refined pathloss construction methodology for CKMs that captures spatial variations in the next section.

\subsection{Traditional Single-antenna CKM Construction Procedure}
\begin{figure}[!t]
	\centering
	\begin{subfigure}{0.48\textwidth} 
		\centering
		\includegraphics[width=\linewidth]{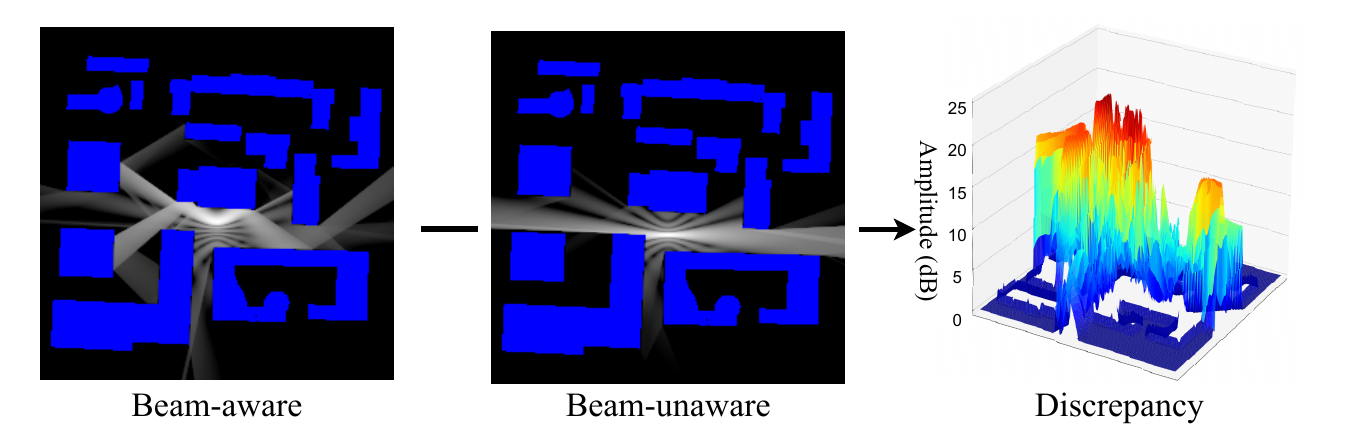} 
		\caption{Comparison of CKMs corresponding to different precoding vectors.}
		\label{fig1:discrpancy}
	\end{subfigure}
	\\
	\vspace{1.5em}
	\centering
	\begin{subfigure}{0.4\textwidth}
		\centering
		\includegraphics[width=\linewidth]{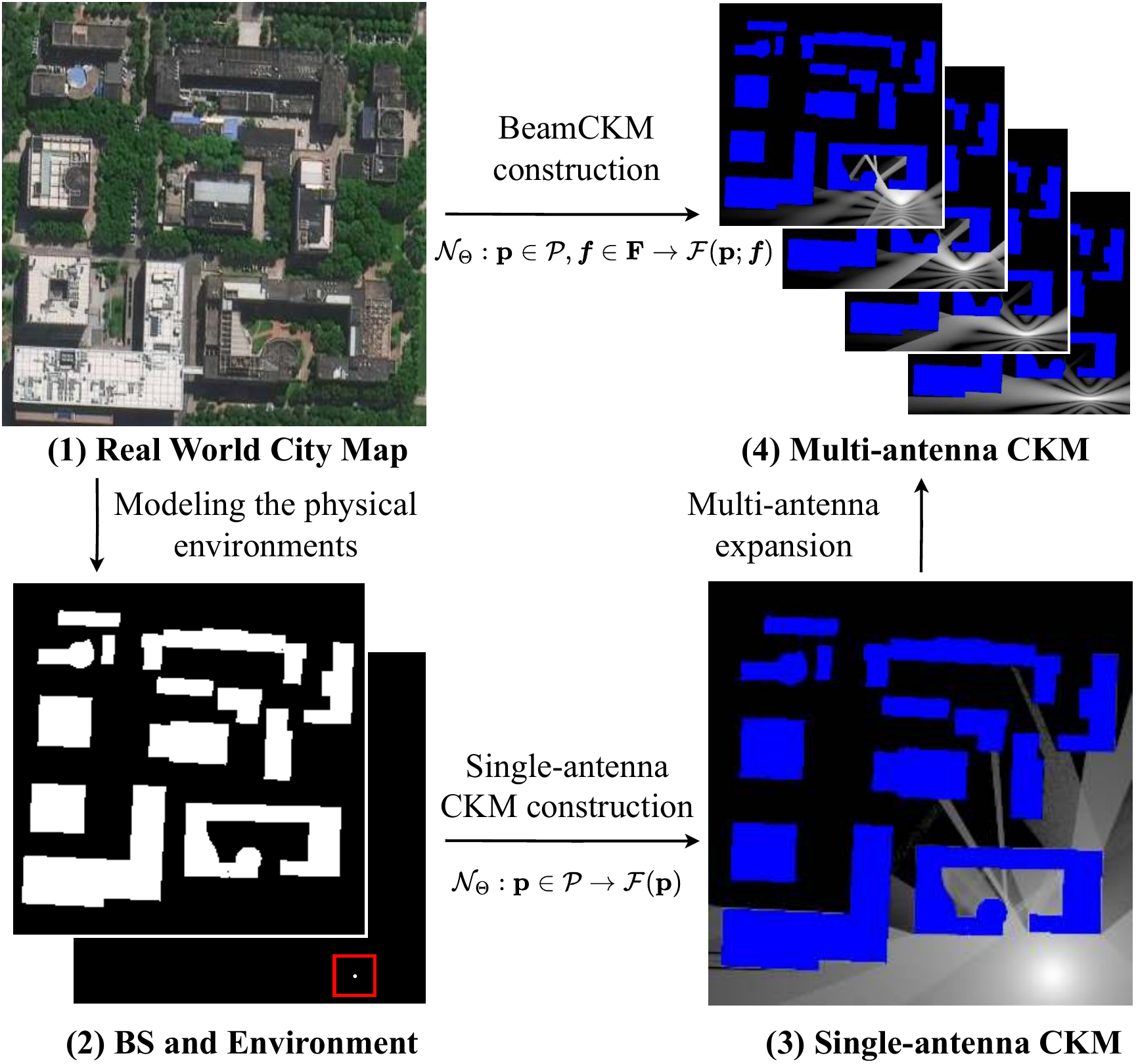}
		\caption{Schematic diagram of real-world modeling and CKM construction.}
		\label{fig1:diagram}
	\end{subfigure}
	\label{fig1}
	\caption{Procedure for constructing CKM in multiple-antenna systems.} 
\end{figure}
The conventional method to estimate channel gain is based on distance-dependent attenuation models. Consider a two-dimensional geographical region $\mathcal{D} \in \mathbb{R}^2$ covered by a transmitter located at point $\mathbf{p}_0 \in \mathbb{R}^2$. Then the pathloss from the BS to another point $\mathbf{p}_i \in \mathcal{D}$ can be expressed in decibels (dB) as:
\begin{equation}
\begin{aligned}
	G_{\mathbf{p}_0\to\mathbf{p}_i}=G_0-10\gamma\log_{10}\left(\|\mathbf{p}_0-\mathbf{p}_i\|\right)\\
	+\alpha_{\mathbf{p}_0\to\mathbf{p}_i}+\beta_{\mathbf{p}_0\to\mathbf{p}_i},
\end{aligned}
\end{equation}
where $G_0$ denotes the path gain at a unit distance, $\gamma$ is the path loss exponent, $\|\cdot\|$ is the Euclidean norm, $\alpha$ accounts for shadow fading, and $\beta$ captures small-scale fading. However, this mathematical model only accounts for distance-dependent attenuation on a large scale and often lacks sufficient accuracy in practical applications. To address this challenge, an effective approach involves constructing CKM in a specific environment.

In detail, we begin by discreting the environment with width $X$ and height $Y$ into a grid of $M = \lceil \frac{X}{\Delta x} \rceil \times \lceil \frac{Y}{\Delta y} \rceil$ pixels, where each pixel represents a physical area of $\Delta x \times \Delta y \, \text{m}^2$. The point set is denoted as $\mathcal{P} = \{\mathbf{p}_i | i=1,2,\ldots,M\}$, where $\mathbf{p}_i = (x_i,y_i)$ represents the coordinates of the $i$-th pixel. The environmental contour is represented by a binary matrix $\boldsymbol{E} \in \{0,1\}^{\lceil \frac{X}{\Delta x} \rceil \times \lceil \frac{Y}{\Delta y} \rceil}$, with 1 denoting buildings. Then the CKM construction process can be formulated as learning a mapping function $\mathcal{F}(\cdot)$ from the environmental contour and BS position to the pathloss values across all points in the target domain, which can be expressed as:
\begin{equation}
	\text{CKM}(\mathcal{P}) \triangleq \mathcal{F}_{\Theta} (\mathcal{P}|\boldsymbol{E}),
\end{equation}
where $\Theta$ denotes the parameters of the mapping function.

\section{Proposed BeamCKM Framework}
In contrast to single-antenna systems, the choice of precoding vector $\boldsymbol{f}$ in multi-antenna configurations significantly affects the received signal strength across the same spatial region \cite{Double}, as illustrated in Fig. \ref{fig1:discrpancy}. Therefore, selecting a suitable precoding vector $\boldsymbol{f}_i$ for the $i$-th user is crucial to maximize the communication rate and minimize inter-user interference \cite{Beamforming}. However, existing studies have primarily focused on optimizing precoding vectors under the assumption of perfect CSI, which typically requires costly channel estimation. To overcome this limitation, a promising alternative is to employ a well-designed codebook and directly generate pathloss CKMs for different codewords. The advantages of this framework can be summarized as follows:
\begin{itemize}
    \item \textit{Unified CKM parameter structure}: It enables structured encoding of channel parameters (e.g., angle, delay, path loss) via discrete beam codebooks. The pathloss-based structure is with unified dimension and similar characteristics, which is convenient for the batch training, inference, and prediction via model/data-driven approach. Thus, the storage consumption and computational overhead are also reduced.
    \item \textit{Incoherent pathloss expression}: It facilitates phase insensitive pathloss estimation while involving the angular beam information. Given the constraints in CKM resolution and positioning accuracy, constructing a phase-coherent CKM is neither feasible nor cost-effective. In other words, the BeamCKM approach substantially enhances both robustness and estimation accuracy.
    \item \textit{Finite codebook-oriented implementation}: It is consistent with the real-world industry deployments with finite beam codebook, including mmWave and THz communication scenarios. The sum-rate and SINR metrics are available under this framework, while the BIM and CAM fail due to the path-specific mapping without phase alignment. Besides, it supports low-overhead CKM construction and update in dynamic environments (e.g., mobile/users) by leveraging a few codewords.
\end{itemize}
In this paper, we utilize the Discrete Fourier Transform (DFT) codebook\footnote{Given the widespread adoption of limited feedback codebooks in practical systems, the DFT codebook serves as the foundation for BeamCKM construction. It should be noted that the proposed framework can accommodate alternative precoding schemes to enhance generality. For practical deployment, the BS only requires training on data generated from its specific codebook.}, defined as:
\begin{equation}
	\begin{aligned}
		\mathbf{F} & = \left[\boldsymbol{f}_0, \boldsymbol{f}_1, \ldots, \boldsymbol{f}_{N_{\text{BS}}-1}\right] \\
		& = \frac{1}{\sqrt{N}}
		\begin{pmatrix}
			1 & 1 & \cdots & 1 \\
			1 & e^{j\frac{2\pi}{N_{\text{BS}}} \cdot 1} & \cdots & e^{j\frac{2\pi}{N_{\text{BS}}} (N_{\text{BS}}-1)} \\
			\vdots & \vdots & \vdots & \vdots \\
			1 & e^{j\frac{2\pi}{N_{\text{BS}}} (N_{\text{BS}}-1)} &  \cdots & e^{j\frac{2\pi}{N_{\text{BS}}} (N_{\text{BS}}-1)^2}
		\end{pmatrix}.
	\end{aligned}
\end{equation}

In an ideal free-space scenario, the DFT codebook enables effective spatial coverage mapping by steering beams in discrete directions, thus facilitating straightforward beam pattern generation in environments without buildings. However, the construction of BeamCKMs under realistic propagation conditions faces challenges due to building-induced multipath reflections and scattering effects, which distort electromagnetic wave propagation and complicate spatial pathloss estimation. Specifically, Fig. \ref{fig1:diagram} illustrates the BeamCKM construction procedure for a given precoding vector $\boldsymbol{f}_j$, where the mapping function is expressed as:
\begin{equation}
\text{CKM}(\mathcal{P}; \boldsymbol{f}_j) \triangleq \mathcal{F}(\mathcal{P}; \boldsymbol{f} = \boldsymbol{f}_j \mid \boldsymbol{E}), \quad j = 0, 1, \ldots, N_{\text{BS}}-1.
\end{equation}
\textit{\textbf{Remark:} It is worth noting that the proposed BeamCKM framework provides a more comprehensive representation compared to existing scalar-based approaches such as CAM and BIM. Our BeamCKM inherently encompasses both spatial and angular channel information by constructing pathloss maps for all codewords in the beamforming codebook. This allows the BeamCKM to not only facilitate beam selection, but also provide detailed channel pathloss information for each potential location under different beams, enabling more sophisticated resource allocation and interference management strategies.}

However, obtaining an analytical expression for the mapping function $\mathcal{F}(\cdot)$ is challenging due to the interactions between electromagnetic waves and environmental structures. Fortunately, with the advancement of DL techniques, it is feasible to approximate this mapping function $\mathcal{F}(\cdot)$ using NNs.
Through supervised learning, the model is trained by minimizing the discrepancy between the predicted BeamCKM and the ground truth measurements. The objective is to minimize the average pathloss estimation error across all locations and codewords, which can be expressed as:
\begin{equation}
\min_{\Theta} \frac{1}{N_{\text{BS}}} \sum_{j=0}^{N_{\text{BS}}-1} \frac{1}{M} \sum_{i=1}^{M} \left| G_{\text{t}}(\mathbf{p}_i; \boldsymbol{f}_j) - \mathcal{F}_{\Theta}(\mathbf{p}_i; \boldsymbol{f}_j \mid \boldsymbol{E}) \right|,
\end{equation}
where $G_{\text{t}}(\mathbf{p}_i; \boldsymbol{f}_j)$ represents the ground truth pathloss at location $\mathbf{p}_i$ under precoding vector $\boldsymbol{f}_j$.

\section{BeamCKM Construction via Environmental Contour}

\begin{figure*}[!t]
	\centering
	\includegraphics[scale=0.33]{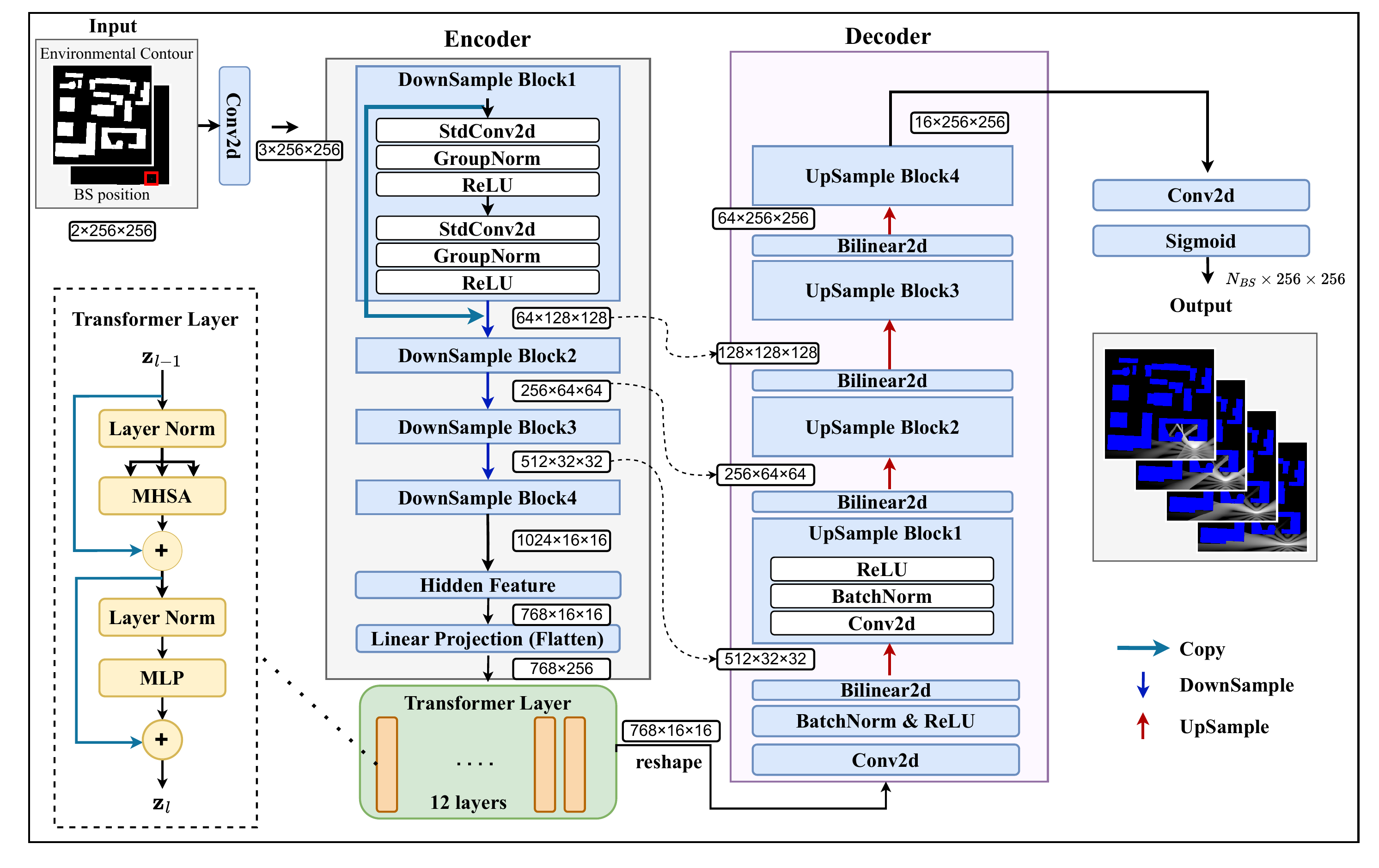} 
	\caption{The detailed structural schematic of CKMTransUNet for BeamCKM construction with environmental contour information.}
	\label{CKMTransUNet}
\end{figure*}

In this section, we develop CKMTransUNet to construct BeamCKM via environmental contour, where the beam patterns and electromagnetic propagation characteristics are fully exploited. The CKMTransUNet architecture first employs CNNs to effectively extract building image features, then utilizes the transformer layer with multihead self-attention (MHSA) and MLP modules to capture long-range dependencies, and finally adopts UNet's structure to integrate contextual information at different resolutions. The overall approach is demonstrated in Fig. \ref{CKMTransUNet}.

\subsection{Hybrid CNN-Transformer Encoder for Feature Extraction}
It is worth noting that BeamCKMs reveal a complex multi-path composition, including specular and diffuse reflections, and multiple bounces. These complex radio frequency phenomena create long-range, non-local dependencies across the entire spatial domain. For instance, a strong reflective surface in one corner of an environment can significantly influence the signal strength received at a distant, NLoS location. To capture these exact types of global contextual relationships, the self-attention mechanism inherent in transformer architectures becomes a natural choice. It computes weighted interactions between pairs of elements in a sequence, enabling the model to weigh the influence of possible reflective surfaces on any given target point. 

The architectural implementation of CKMTransUNet shares fundamental similarities with standard ViTs. To establish a technical foundation, we first delineate the canonical ViT architecture before highlighting the distinctive characteristics of CKMTransUNet. 
The fundamental procedure of ViT is as follows. Considering the $C \times H \times W$ dimensional input, the image is first partitioned into $P \times P$ patches, converting the $C \times H \times W$ input into $C \times N \times P \times P$ patch blocks, where $C$ is the input channel and $N = \frac{HW}{P^2}$ is the number of image patches. Next, these blocks are flattened to $\mathbf{x}_p^1, \mathbf{x}_p^2, ..., \mathbf{x}_p^N$, where $\mathbf{x}_p^i \in \mathbb{R}^{1\times P^2 C}, \forall i = 1,2,...,N$. Then patches $\mathbf{x}_p^i, \forall i$ are processed by the patch embedding module, which is given by:
\begin{equation}
	\mathbf{z}_0=[\mathbf{x}_p^1\mathbf{E};\mathbf{~x}_p^2\mathbf{E};\cdots;\mathbf{~x}_p^N\mathbf{E}]+\mathbf{E}_{\text{pos}},
\end{equation}
where $\mathbf{E}\in\mathbb{R}^{P^2C \times D}$ denotes the patch embedding projection, and $\mathbf{E}_{\text{pos}}\in\mathbb{R}^{N\times D}$ denotes the position embedding, which is used to maintain the positional information.
The embedded sequence $\mathbf{z}_0$ subsequently propagates through transformer modules that capture contextual relationships via attention mechanisms. The process of the $l$-th layer can be written as follows:
\begin{equation}
	\begin{aligned}
		& \mathbf{z}_l^{\prime}=\mathrm{MHSA}(\mathrm{LN}(\mathbf{z}_{l-1}))+\mathbf{z}_{l-1}, \\
		& \mathbf{z}_l=\mathrm{MLP}(\mathrm{LN}(\mathbf{z}_l^{\prime}))+\mathbf{z}_l^{\prime},
	\end{aligned}
\end{equation}
where $\mathrm{LN}(\cdot)$ denotes the layer normalization. The structure of a transformer layer is demonstrated in the left panel of Fig. \ref{CKMTransUNet}.

In contrast to a pure transformer encoder, CKMTransUNet adopts a hybrid CNN and transformer architecture. The CNN backbone first performs hierarchical downsampling with ResNet rather than using the ViT patches sized $P \times P$. The intermediate layer dimensions include $\{64,256,512,1024\}$. Then the hidden feature is flattened into a sequence of linear projection through a $1 \times 1$ convolution, after which the vision problem is transformed into a seq2seq problem. After patch embedding, the embedded features are propagated to the transformer layer. \cite{TransUnet} discovers that the hybrid CNN-transformer encoder performs better than using a pure ViT. This can be attributed to the CNN-based downsampling process, which demonstrates a superior capability in extracting information from diverse spatial locations within an image compared to direct transformation via patch embedding into linear layers.

\subsection{Cascaded UNet-structured Upsampler for Fine-grained Reconstruction}
The decoder employs a cascaded upsampling architecture to progressively reconstruct the high-resolution output from the encoded features, transforming the spatial dimensions from $\frac{H}{2^{S}} \times \frac{W}{2^{S}}$ back to $H \times W$, where $S$ is the number of downsampling stages. Each upsampling module integrates an upsampling operator, convolutional layers, and ReLU activation.
A critical component is the incorporation of skip connections that link the encoder and decoder. This mechanism allows high-resolution, low-level feature maps $\mathbf{F}_{i}$ from the encoder to be combined with the upsampled features $\mathbf{D}_{i}$ from the decoder, which is formally described by:
\begin{equation}
	\mathbf{D}_{i} = \text{Upsample}(\text{Concat}(\mathbf{D}_{i+1}, \mathbf{F}_i)).
\end{equation}
This process ensures that the spatial details captured by the early encoder layers are directly utilized by the decoder for precise construction, effectively compensating for information loss during downsampling.
The final stage involves a convolutional layer and a sigmoid activation to adjust the dimension and generate the pixel-wise prediction. Collectively, the synergistic integration of the encoder and decoder, facilitated by these skip connections, enables multi-resolution feature aggregation and forms a distinctive U-shaped architecture.

\subsection{Composite Loss Design}
\begin{figure}[!t]
	\begin{subfigure}{0.48\textwidth}
		\centering
		\includegraphics[width=\linewidth]{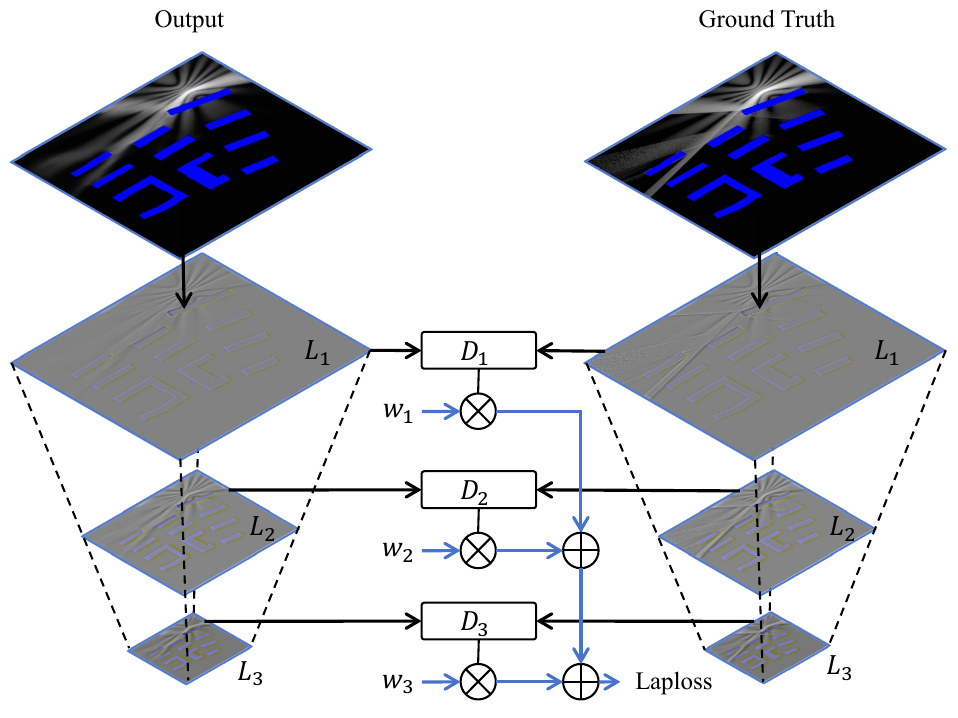}
		\caption{Schematic diagram of Laplacian loss.}
		\label{loss:lap}
	\end{subfigure}

	\vspace{1em}

	\begin{subfigure}{0.48\textwidth} 
		\centering
		\includegraphics[width=\linewidth]{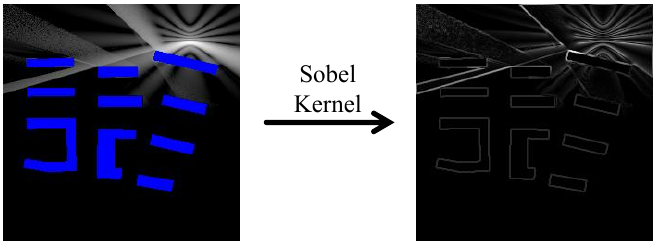} 
		\caption{The comparsion of initial maps and edge-extracted maps via Sobel operator.}
		\label{loss:sobel}
	\end{subfigure}
	\label{DesignLoss}
	\caption{Schematic diagram of Laplacian loss and edge loss.} 
\end{figure}

In the construction of BeamCKM, the electromagnetic wave intensity distribution in $\mathcal{P}$ manifests pronounced edge discontinuities induced by specular reflections and beamforming effects. To better capture these features using NNs, the proposed approach employs a designed composite loss architecture to guide the optimization process. As shown in Eq. (\ref{eq:total_loss}), the total loss combines four distinct components with chosen weighting coefficients:
\begin{equation}
\begin{split}
    L_{\text{total}} = 
	\underbrace{\lambda_1 L_{2}}_{\text{pixel accuracy}} 
	&+ \underbrace{\lambda_2 L_{\text{lap}}}_{\text{multi-scale detail}} \\
    + &\underbrace{\lambda_3 L_{\text{edge}}}_{\text{structural preservation}} + \lambda_4 L_{\text{ssim}},
\end{split}
\label{eq:total_loss}
\end{equation}
where $\{\lambda_1, \lambda_2, \lambda_3, \lambda_4\} = \{1, 0.01, 0.005, 0.001\}$ denote the relative importance weights for each component, determined through empirical validation. The details of each loss component are elaborated as follows.

\subsubsection{L2 (Mean Squared Error) Loss}
The foundation of our loss function is the standard L2 norm, which ensures basic pixel-level accuracy:
\begin{equation}
	\label{eq:l2_loss}
	L_{2} = \frac{1}{M}\sum_{i=1}^{M}|G_{\text{pred}}^{(i)} - G_{\text{true}}^{(i)}|^2,
\end{equation}
where $M = H \times W$ is the total number of pixels in the image of height $H$ and width $W$. This component provides a strong baseline for image reconstruction quality.

\subsubsection{Laplacian Pyramid Loss}
To preserve multi-scale structural fidelity in image generation tasks, we implement a Laplacian pyramid decomposition approach that captures structural differences across multiple spatial frequencies. The pyramid construction process begins with building a Gaussian pyramid through iterative downsampling, where each level applies a depthwise convolution with a fixed $5 \times 5$ Gaussian kernel followed by stride-2 subsampling.
The Laplacian pyramid is then formed by computing the difference between each Gaussian level and the upsampled version of the next higher level, which can be expressed as $L_i = G_i - \text{upsample}(G_{i+1})$ for $i < \text{levels}-1$, with the top level consisting of the lowest-resolution Gaussian. The upsampling operation utilizes bilinear interpolation followed by Gaussian blurring to maintain frequency consistency between pyramid levels.
The multi-scale comparison is performed using a weighted L1 norm between corresponding levels of the pyramids constructed from predicted and target images, which is demonstrated in Fig. \ref{loss:lap}:
\begin{equation}
	L_{\text{lap}} = \sum_{l=0}^{\text{levels}-1} w_l \| L_l(G_{\text{pred}}) - L_l(G_{\text{true}}) \|_1,
\end{equation}
where $w_l = 0.5^l$ represents geometrically decaying weights that emphasize finer details at higher pyramid levels. This approach ensures that texture details are preserved across different spatial frequencies.

\subsubsection{Edge-Aware Loss}
For enhanced structural preservation, we derive an edge-aware loss through Sobel gradient operators. First, compute horizontal ($G_x$) and vertical ($G_y$) gradients through convolution with Sobel kernels.
Then calculating edge strength maps with numerical stability constant $\epsilon = 10^{-6}$ yields:
\begin{equation}
	E = \sqrt{G_x^2 + G_y^2 + \epsilon}.
\end{equation}
The final edge loss, which combines L1 differences, is given by:
\begin{equation}
	\label{eq:edge_loss}
	L_{\text{edge}} =  \|E_{\text{pred}} - E_{\text{true}}\|_1.
\end{equation}
This loss emphasizes the preservation of sharp transitions and structural details, as illustrated in Fig. \ref{loss:sobel}.

\subsubsection{SSIM (Structural Similarity Index Measure) Loss}
For enhanced preservation of structural similarity, we implement a SSIM loss through Gaussian-weighted window comparison. Let $x = G_{\text{pred}}$ and $y = G_{\text{true}}$ denote the predicted and ground truth images, respectively. First, we compute local means ($\mu_x$, $\mu_y$), variances ($\sigma_x^2$, $\sigma_y^2$), and cross-covariance ($\sigma_{xy}$) using Gaussian convolution kernels with standard deviation $\sigma=1.5$ and window size $11\times11$. The SSIM index combines luminance, contrast, and structure comparisons with stability constants $C_1,C_2$:
\begin{equation}
	\label{SSIM_expression}
	\text{SSIM}(x, y) = \frac{(2\mu_x\mu_y + C_1)(2\sigma_{xy} + C_2)}{(\mu_x^2 + \mu_y^2 + C_1)(\sigma_x^2 + \sigma_y^2 + C_2)}.
\end{equation}
The final SSIM loss aggregates the structural dissimilarity across spatial dimensions:
\begin{equation}
	\label{eq:ssim_loss}
	L_{\text{ssim}} = 1 - \text{SSIM}(G_{\text{pred}}, G_{\text{true}}).
\end{equation}

\section{BeamCKM Construction with Sparse Observations via Multi-modal Learning}
\begin{figure*}[!h]
	\centering
	\hspace*{0cm}
	\includegraphics[scale=0.50]{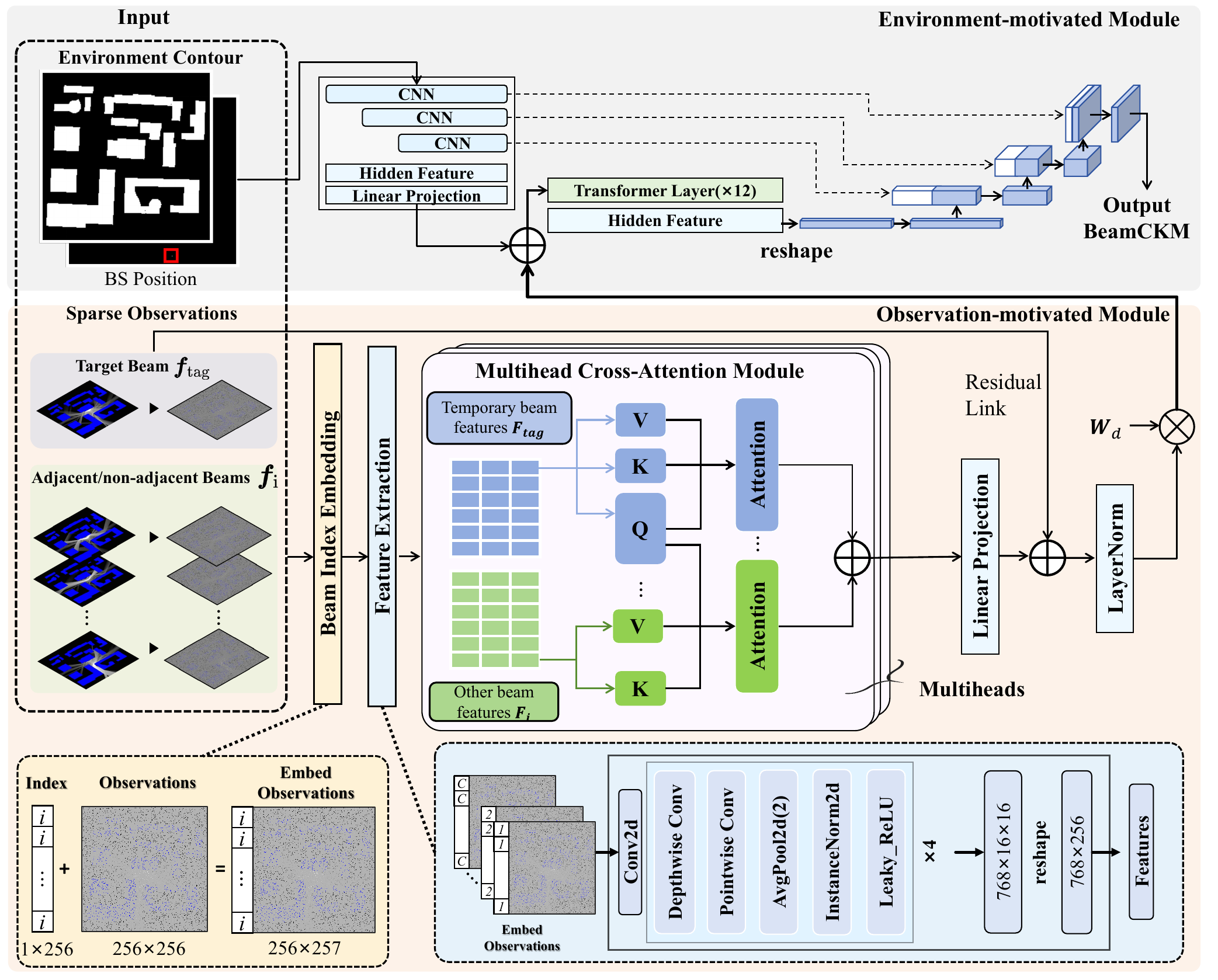} 
	\caption{Schematic of multi-modal BeamCKM construction incorporating environmental contour and sparse observations.}
	\label{cross_attention}
\end{figure*}

Considering the time-varying nature of the environment $\boldsymbol{E}$, the BS can leverage sparse observations $\mathcal{S}$ obtained from real-time sensor networks or crowdsourcing mechanisms \cite{crowdsourcing} to enhance the construction accuracy. In other words, these sparse observations can be utilized in a data-driven framework to guide the BeamCKM construction.
Unlike single-antenna systems, where observations are generally beam-independent, in multi-antenna systems, the selection of different codewords $\boldsymbol{f}$ results in distinct spatial sampling patterns. Consequently, the sparse observations collected under different beams exhibit unique spatial distributions and channel characteristics.
This section investigates how observations from multiple beams can be effectively integrated to enhance BeamCKM construction accuracy.

Given the modality-rich nature of multi-beam observations, we formulate the construction task as a multi-modal learning problem, where each beam's data provides a distinct yet complementary information modality.
On one hand, beams steered in similar directions often illuminate overlapping spatial regions. This creates informational redundancy, which helps suppress noise. On the other hand, beams pointing in different directions give complementary coverage and illuminate more of the environment together than any alone. Utilizing the redundancy and complementarity can enhance BeamCKM precision, as described in these aspects:
\begin{itemize}
    \item \textit{Adjacent Codeword Synergy: Beamforming codewords with small angular separation ($\| \theta_i-\theta_j \| \leq \delta$) typically sample spatially contiguous regions, inducing high correlation between observations $\mathcal{S}_i$ and $\mathcal{S}_j$. This local smoothness prior enables reliable information within coverage gaps.}
    \item \textit{Non-Local Information Transfer: Even beams with non-overlapping coverage can exhibit statistical dependencies due to shared environments. By exploiting these non-local relationships through cross-attention mechanisms, the multi-modal scheme effectively transfers contextual information from different modalities across beams, enhancing construction accuracy and robustness.}
\end{itemize}
It is worth noting that the construction outputs change from all beams to a specific target beam. Accordingly, we name this multi-modal multi-beam CKM network under observable conditions as $\text{M}^3$ChanNet. The overall procedure is demonstrated in Fig. \ref{cross_attention} and the cross-attention mechanism constitutes the basis of our BeamCKM reconstruction framework. This mechanism operates through the BeamAttentionEncoder module, which is introduced as follows.

\subsection{Initial Processing}
The beam feature extraction pipeline employs a hierarchical processing sequence, which includes:
	\begin{enumerate}
		\item \textit{Data Identification and Filling}: Unsampled points are identified and filled with $-1$:
		\begin{equation}
			\mathcal{S}_{i}^{\prime} = \mathbb{I}_{\text{sampled}}(\mathcal{S}_i) \odot \mathcal{S}_i + \mathbb{I}_{\text{unsampled}}(\mathcal{S}_i) \odot (-1), \forall i,
		\end{equation}
		where $\mathbb{I}_{\text{sampled}}(\cdot)$ and $\mathbb{I}_{\text{unsampled}}(\cdot)$ are indicator functions for valid and unsampled data, respectively, and $\odot$ denotes Hadamard product. This step ensures that only valid measurements contribute to subsequent feature extraction.
		\item \textit{Beam Index Embedding}: Codeword indices are integrated via spatial concatenation:
		\begin{equation}
			\mathcal{S}_{i}^{\text{embed}} = \text{Concat}\left( \mathcal{S}_{i}^{\prime}, \mathcal{I} \right)
		\end{equation}
		where $\mathcal{I} \in \mathbb{R}^{1 \times W}$ is constructed by repeating the beam index $i$ to form beam index embeddings, and $\text{Concat}(\cdot)$ denotes the concatenation operator. This explicitly marks the specific beam associated with each observation. This marking enables the system to discern whether other observations originate from adjacent beams or from those with substantially different indices.
		It is worth noting that while alternative encodings (e.g., one-hot or learnable embeddings) are feasible, scalar repetition is adopted for its operational simplicity, effectively providing the necessary beam identity information.
		\item \textit{Low-complexity Feature Extracting}: A depthwise separable convolutional network is applied to achieve efficient feature extraction:
		\begin{equation}
			\mathbf{F}_{i}^{\text{feat}} = \mathcal{N}_{\text{DS}} \left( \mathcal{S}_i^{\text{embed}} \right) \in \mathbb{R}^{hw \times D},
			\label{DS}
		\end{equation}
		where $\mathcal{N}_{\text{DS}}(\cdot)$ denotes the feature extraction module in Fig. \ref{cross_attention}, which consists of depthwise separable convolutional layers followed by batch normalization and LeakyReLU activation. Here, $hw$ represents the flattened dimensions, and $D$ is the channel dimension. This design significantly reduces computational complexity while effectively capturing spatial features from sparse observations.
	\end{enumerate}
\subsection{Cross Attention Mechanism}
The core of the BeamAttentionEncoder module is a multi-head cross-attention mechanism, which integrates different beam features to enhance the target feature representation. The implementation involves the following steps:
\begin{enumerate}
    \item \textit{Multi-Head Attention Scheme}: 
    The attention mechanism computes weighted interactions between target beam features $\mathbf{F}_{\text{tag}}$ and other beam features $\{\mathbf{F}_i\}_{i=1}^C$. This is achieved through $H$ parallel attention heads:
	\begin{equation}
		\begin{split}
			\text{MultiHead}&(\mathbf{Q}_{\text{tag}},\mathbf{K}_i,\mathbf{V}_i) = \\
			&\quad \text{Concat}(\text{head}_{1,i}, \dots, \text{head}_{H,i})\mathbf{W}^O,
		\end{split}
	\end{equation}
    where $\mathbf{W}^O \in \mathbb{R}^{256 \times 256}$ is the output projection matrix.
    The multi-head attention mechanism projects the input into $H$ lower-dimensional subspaces. This provides the model with multiple representation subspaces to capture complementary modal characteristics.
    \item \textit{Single-Head Attention Branch}: 
    Each attention head computes a standard attention mechanism:
    \begin{equation}
		\begin{split}
			\text{head}_i^h &= \text{CrossAttn}(\mathbf{Q}_{h,\text{tag}}, \mathbf{K}_{h,i}, \mathbf{V}_{h,i}) \\
			&= \text{softmax}\left(\frac{\mathbf{Q}_{h,\text{tag}} {\mathbf{K}_{h,i}}^T}{\sqrt{d}} \right) \mathbf{V}_{h,i},
		\end{split}
    \end{equation}
    with queries $\mathbf{Q}_{h,\text{tag}}= \mathbf{F}_{\text{tag}}^{\text{feat}}\mathbf{W}^Q_{h,\text{tag}}$, keys $\mathbf{K}_{h,i} = \mathbf{F}_{i}^{\text{feat}}\mathbf{W}^K_{h,i}$, and values $\mathbf{V}_{h,i} = \mathbf{F}_{i}^{\text{feat}}\mathbf{W}_{h,i}^V$ for head $h$. Here, $\sqrt{d}$ is the scaling factor that stabilizes the training by preventing the products from growing large in magnitude.
	If the target beam observations can not be obtained, the query $\mathbf{Q}_{h,\text{tag}}$ can be generated from the environmental contour $\boldsymbol{E}$ features via feature extraction module $\mathcal{N}_{\text{DS}}(\cdot)$ in Eq. (\ref{DS}).
    \item \textit{Aggregation across Beams}:
    The outputs from all beams are aggregated to form the combined feature representation:
	\begin{equation}
		\mathbf{F}_{\text{multi}} = \sum_{i=1}^{C} \text{MultiHead}(\mathbf{Q}_{\text{tag}}, \mathbf{K}_i, \mathbf{V}_i).
	\end{equation}
    \item \textit{Residual Enhancement}: 
    The multi-head output is combined with the original reference features via residual connection and layer normalization:
    \begin{equation}
        \mathbf{F}_{\text{attn}} = \text{LN}\left( \mathbf{F}_{\text{multi}} + \mathbf{F}^{\text{feat}}_{\text{tag}} \right).
    \end{equation}
\end{enumerate}

\subsection{High-fidelity Signal Retention}
To ensure the integrity of original details and adaptively handle sparse observations, the scheme incorporates two key components:
\begin{enumerate}
    \item \textit{Residual Connection}: The output of the attention mechanism is fused with the original beam observations through a residual pathway to preserve fine-grained details:
    \begin{equation}
        \mathbf{F}_{\text{beam}} = \text{LN} \left( \mathbf{FC}(\mathbf{F}_{\text{attn}}) + \mathbf{B}(\mathcal{S}_{\text{tag}}) \right),
    \end{equation}
    where $\mathbf{FC}(\cdot)$ denotes a linear projection that aligns the dimension of $\mathbf{F}_{\text{attn}}$, and $\mathbf{B}(\cdot)$ denotes a bilinear resampling operation that resizes the original beam observation $\mathcal{S}_{\text{tag}}$ to match the spatial size of $\mathbf{F}_{\text{attn}}$.

	\item \textit{Observations Density-Weighting}: A spatial weight is generated based on local observation density to enhance robustness in undersampled regions. The density map $\rho \in \mathbb{R}^{B \times H \times W}$ is first computed as:
    \begin{equation}
        \rho = \frac{1}{C} \sum_{i=1}^C \mathbb{I}(\mathcal{S}^{\prime}_i \neq -1),
    \end{equation}
    which averages the binary observation masks across all beams. This density map is then processed by a small convolutional network, followed by a reshaping operation to ensure dimensional compatibility with the beam features. The final density weights are produced as:
    \begin{equation}
        \textbf{W}_d = \text{reshape} \left( \sigma \left( \text{Avg} \left( \mathbf{Conv}_{3\times3} \left( \rho \right) \right) \right) \right),
    \end{equation}
    where $\sigma(\cdot)$ is the sigmoid function. The beam features are adaptively scaled by $\textbf{W}_d$ to emphasize regions with reliable observations. Finally, the output features are computed as $\mathbf{F}_{\text{out}} = \mathbf{F}_{\text{beam}} \odot \textbf{W}_d$.
\end{enumerate}
After the BeamAttentionEncoder, the results are finally added with the initial transformer input features together into the transformer layer. The whole procedure is summarized in Algorithm 1.

\begin{algorithm}[!t]
	\caption{Observation-motivated multi-modal BeamCKM construction ($\text{M}^3$ChanNet)}
	\label{alg:beam_attention_encoder}
	\begin{algorithmic}[1]
		\Require Sparse observations $\mathcal{S} \in \mathbb{R}^{C \times H \times W}$, target beam index, and environmental profile.
		
		\State \textit{\textbf{Initial Processing:}}
		\State $\mathcal{S}^{\prime} = \mathbb{I}_{\text{sampled}}(\mathcal{S}) \odot \mathcal{S} + \mathbb{I}_{\text{unsampled}}(\mathcal{S}) \odot (-1)$
		\State $\mathcal{S}^{\text{embed}} = \text{Concat}(\mathcal{S}^{\prime}, \mathcal{I})$
		\State $\mathbf{F}^{\text{feat}} = \mathcal{N}_{\text{DSConv}}(\mathcal{S}^{\text{embed}})$
		\State \textit{\textbf{Multi-Head Cross-Attention:}}
		\State $\mathbf{F}_{\text{multi}} = \mathbf{0}$
		\For{$i = 1$ to $C$}
		\For{$h = 1$ to $H$}
			\State $\mathbf{Q}_{h,\text{tag}} = \mathbf{F}_{\text{tag}}^{\text{feat}}\mathbf{W}^Q_h$;
			\State $\mathbf{K}_{h,i} = \mathbf{F}_{i}^{\text{feat}}\mathbf{W}^K_h$;
			\State $\mathbf{V}_{h,i} = \mathbf{F}_{i}^{\text{feat}}\mathbf{W}^V_h$;
			\State $\text{head}_{h,i} = \text{softmax}\left(\frac{\mathbf{Q}_{h,\text{tag}}\mathbf{K}_{h,i}^T}{\sqrt{d}}\right)\mathbf{V}_{h,i}$
		\EndFor
			\State $\mathbf{F}_{\text{multi}} \mathrel{+}= \text{Concat}(\text{head}_{1,i}, \dots, \text{head}_{H,i})\mathbf{W}^O$
		\EndFor
		\State $\mathbf{F}_{\text{attn}} = \text{LN}(\mathbf{F}_{\text{multi}} + \mathbf{F}_{\text{tag}}^{\text{feat}})$
		
		\State \textit{\textbf{High-Fidelity Signal Retention:}}
		\State $\mathbf{F}_{\text{beam}} = \text{LN}(\mathbf{FC}(\mathbf{F}_{\text{attn}}) + \mathbf{B}(\mathcal{S}_{\text{tag}}))$
		\State $\rho = \frac{1}{C}\sum_{i=1}^C \mathbb{I}(\mathcal{S}^{\prime}_i \neq -1)$
		\State $\textbf{W}_d = \text{reshape}(\sigma(\mathbf{Avg}(\mathbf{Conv}_{3\times3}(\rho))))$
		\State $\mathbf{F}_{\text{out}} = \mathbf{F}_{\text{beam}} \odot \textbf{W}_d$
		
		\State \Return $\mathbf{F}_{\text{out}}$
	\end{algorithmic}
\end{algorithm}

\section{Simulation Results}
This section evaluates the performance of our proposed CKMTransUNet and $\text{M}^3$ChanNet architectures through comprehensive experiments. We begin by introducing the experimental setup, including the dataset and parameter configurations. Subsequently, we compare our CKMTransUNet and $\text{M}^3$ChanNet against SOTA methods to assess their accuracy in BeamCKM construction. Further, we present ablation studies to verify the contribution of each module within the proposed approach. Finally, we compare the robustness of different methods under noisy environmental contours.

\subsection{BeamCKM Dataset}
The BeamCKM dataset comprises 100 geo-referenced urban maps sampled from representative Chinese cities, with building contours extracted from OpenStreetMap \cite{Openstreetmap}. For each geographical map, 100 BS locations with ULA antennas are randomly deployed at a height of 1.5m\footnote{The goal of this study is to investigate the construction of CKM under various precoding vectors. When BSs are deployed at different heights, the methodology illustrated in Sionna can be employed to generate the corresponding dataset, thereby enabling BeamCKM construction for BSs at arbitrary altitudes. }, operating at 3GHz carrier frequency. The dataset generation employs the Sionna ray-tracing toolkit with $10^9$ rays per simulation and realistic electromagnetic parameters following ITU-R standards to ensure physical accuracy. 

Each scenario is simulated using all codewords from the DFT codebook as precoding vectors with $N_{BS}=8$, generating comprehensive BeamCKMs. The output data is structured as $256\times256$ pixel images where each pixel corresponds to 1m² in the physical environment. Pathloss values undergo normalization processing with a 40dB dynamic range, enabling reconstruction of received signal strength and SINR for practical beamforming optimization. This dataset facilitates the development of digital twins and provides a foundation for studying codebook-specific channel characteristics across diverse urban environments.

\subsection{Parameter Setting}

A consistent set of experimental configurations is adopted across all studies in this paper. All models are trained on an NVIDIA RTX 5090 GPU equipped with 32 GB of memory. The AdamW optimizer is utilized with an initial learning rate of $10^{-4}$, scheduled to decay by a factor of 0.1 every 20 epochs. The dataset is divided into training, validation, and test subsets following ratio 8:1:1. Evaluations are conducted on random observation subsets comprising \{0.5\%,1\%,2\%,3\%,4\%,5\%\} of the total pixels. Our models are built upon PyTorch, configured with a patch size of 16 and a transformer architecture of 12 layers. Execution is optimized using the FlashAttention and Accelerate toolkits for lower-precision calculation, effectively decreasing training duration and GPU memory consumption. To assess the model's performance, we conduct extensive comparisons with SOTA methods, which are implemented as follows:
\begin{itemize}
	\item \textbf{\textit{Kriging}}: Kriging is a classical geostatistical interpolation technique that utilizes spatial autocorrelation to estimate unknown values at unobserved locations, which has been widely adopted for spatial prediction and mapping tasks.
	\item \textbf{\textit{UNet}}: UNet represents one of the most prominent CNN architectures, which has been extensively employed across various CV tasks, including image segmentation, prediction, regression, and classification problems.
	\item \textbf{\textit{RadioWNet}}: RadioWNet is first proposed in \cite{RadioUnet}. The proposed architecture employs a two-stage cascaded UNet approach, where the output from the first UNet is concatenated with the original input to serve as the input for the second UNet. While maintaining a relatively simple structure, this cascaded design demonstrates remarkable capability in CKM construction.
	\item \textbf{\textit{RME-GAN}}: RME-GAN is put forward in \cite{GAN}. The model employs a generative adversarial network (GAN) approach, where a generator synthesizes CKM while a discriminator evaluates their authenticity. This model performs well when observations exist.
\end{itemize}

\noindent To evaluate the performance of different methods, the following metrics are adopted. The root mean square error (RMSE) reflects the absolute deviation between predicted values and ground truth. The normalized mean square error (NMSE) normalizes the error using the variance of the ground truth data. The peak signal-to-noise ratio (PSNR) measures the ratio between the maximum possible signal power and the power of corrupting noise in image reconstruction. The structural similarity index (SSIM) evaluates the perceptual similarity between images by comparing their luminance, contrast, and structural information.

\subsection{Simulation without Observations}
\begin{figure*}[!htbp]
    \centering
    \begin{subfigure}[t]{0.16\textwidth}
        \includegraphics[width=\textwidth]{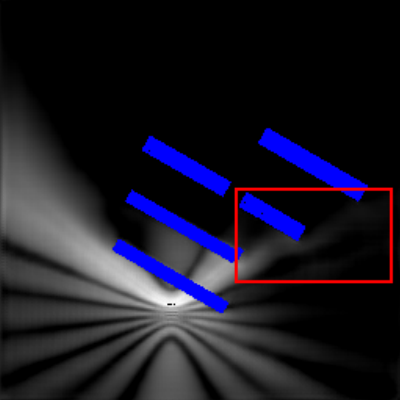}
        \label{fig:rme-gan_a}
    \end{subfigure}
    \hfill
    \begin{subfigure}[t]{0.16\textwidth}
        \includegraphics[width=\textwidth]{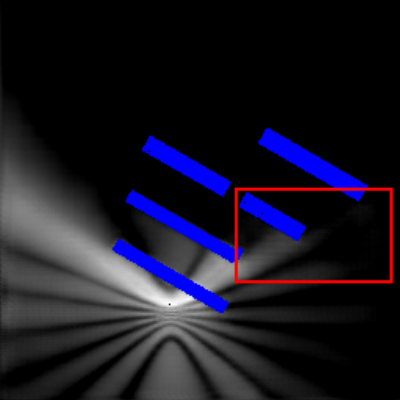}
        \label{fig:unet_b}
    \end{subfigure}
    \hfill
    \begin{subfigure}[t]{0.16\textwidth}
        \includegraphics[width=\textwidth]{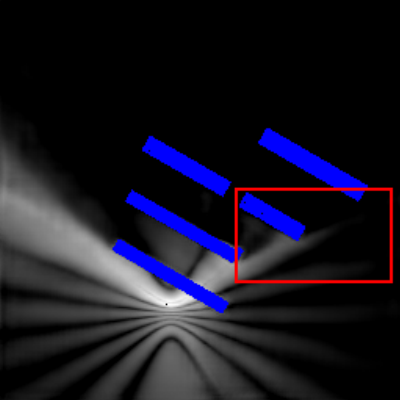}
        \label{fig:radiownet_c}
    \end{subfigure}
    \hfill
    \begin{subfigure}[t]{0.16\textwidth}
        \includegraphics[width=\textwidth]{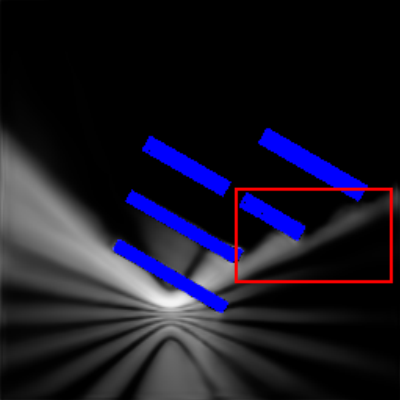}
        \label{fig:transunet_d}
    \end{subfigure}
    \hfill
    \begin{subfigure}[t]{0.16\textwidth}
        \includegraphics[width=\textwidth]{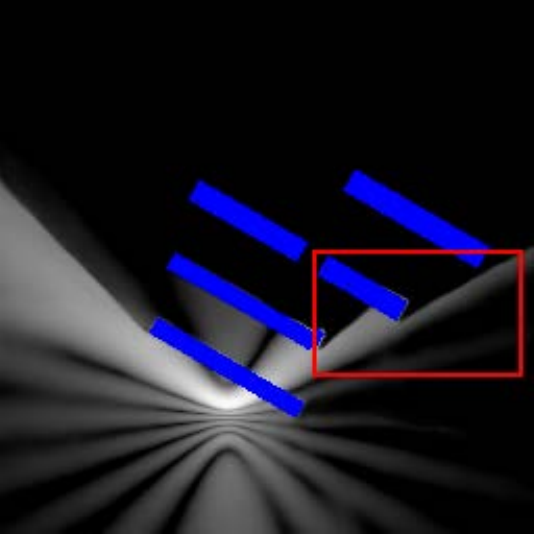}
        \label{fig:transunet_e}
    \end{subfigure}
    \hfill
    \begin{subfigure}[t]{0.16\textwidth}
        \includegraphics[width=\textwidth]{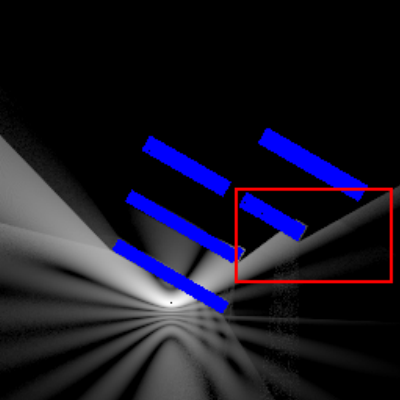}
        \label{fig:groundtruth_f}
    \end{subfigure}
    
    \vspace{0.5cm}
    
    \begin{subfigure}[t]{0.16\textwidth}
        \includegraphics[width=\textwidth]{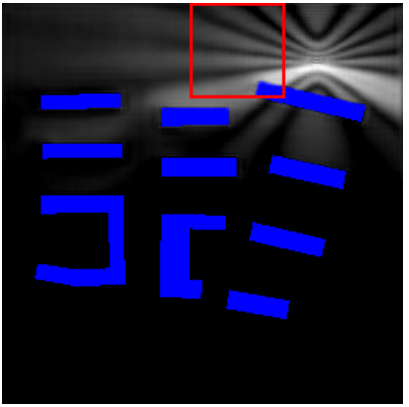}
        \caption{RME-GAN}
        \label{fig:rme-gan_a2}
    \end{subfigure}
    \hfill
    \begin{subfigure}[t]{0.16\textwidth}
        \includegraphics[width=\textwidth]{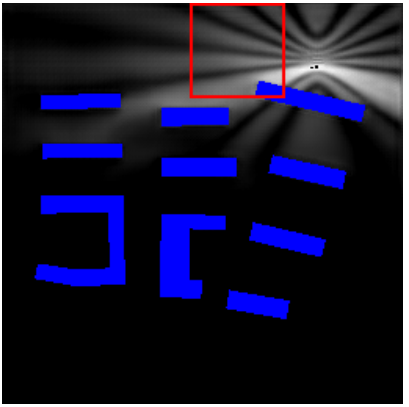}
        \caption{UNet}
        \label{fig:unet_b2}
    \end{subfigure}
    \hfill
    \begin{subfigure}[t]{0.16\textwidth}
        \includegraphics[width=\textwidth]{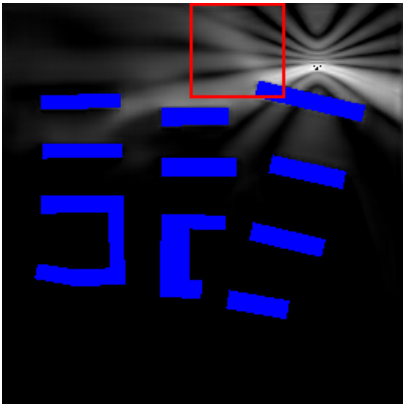}
        \caption{RadioWNet}
        \label{fig:radiownet_c2}
    \end{subfigure}
    \hfill
    \begin{subfigure}[t]{0.16\textwidth}
        \includegraphics[width=\textwidth]{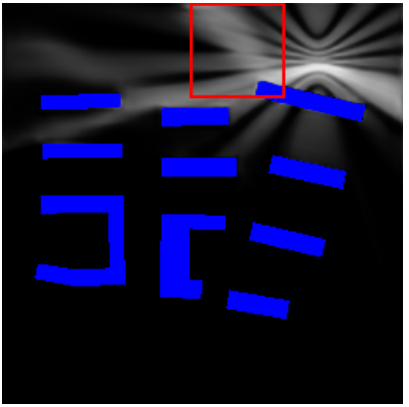}
        \caption{CKMTransUNet (w/o observation)}
        \label{fig:transunet_d2}
    \end{subfigure}
    \hfill
    \begin{subfigure}[t]{0.16\textwidth}
        \includegraphics[width=\textwidth]{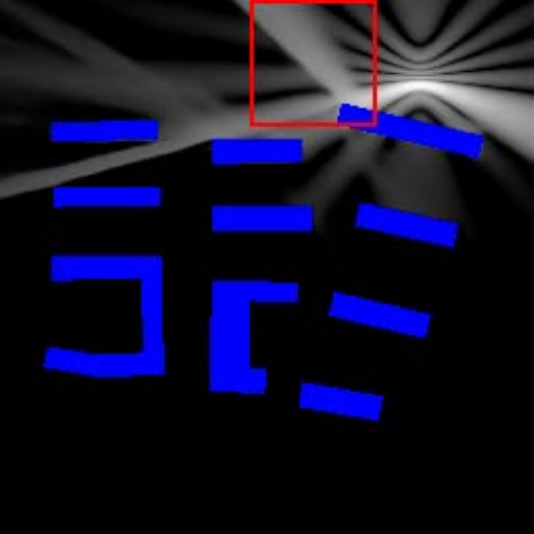}
        \caption{$\text{M}^3$ChanNet (w/ 1\% observation)}
        \label{fig:transunet_e2}
    \end{subfigure}
    \hfill
    \begin{subfigure}[t]{0.16\textwidth}
        \includegraphics[width=\textwidth]{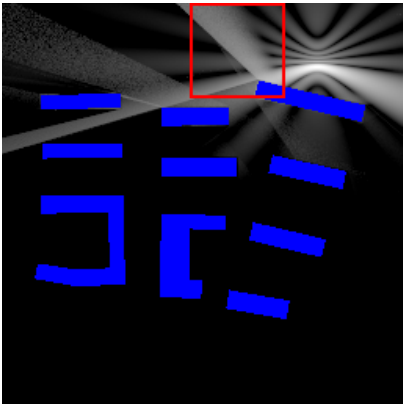}
        \caption{Ground truth}
        \label{fig:groundtruth_f2}
    \end{subfigure}
    
    \caption{Visualization of BeamCKM construction results under specific beamforming codewords.}
    \label{radio_map_comparison_fig}
\end{figure*}

\begin{figure}[!h]
	\centering
	\includegraphics[scale=0.45]{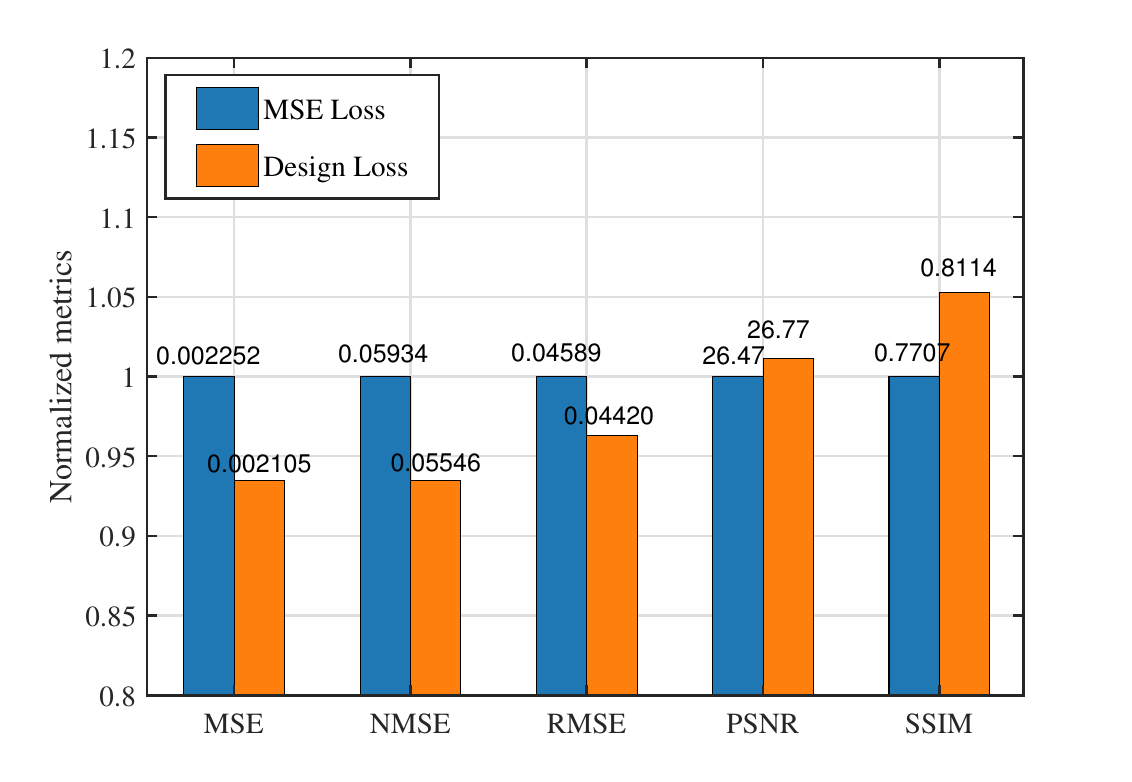}
	\caption{Metrics with different loss functions for CKMTransUNet.}
	\label{losscompare}
\end{figure}

We first compare the results with CKMTransUNet under the no-observation scenario. As shown in Table \ref{Compare_without_observation}, CKMTransUNet consistently outperforms all baseline methods across every evaluation metric. Notably, it achieves a 33.4\% improvement in MSE over RadioWNet, which highlights its remarkable capabilities. The results presented in Fig. \ref{radio_map_comparison_fig} show the details and reveal that CKMTransUNet achieves superior performance compared to other methods.
We further validate the effectiveness of the proposed loss architecture. While keeping all other experimental variables unchanged, we modify only the loss function during training. As shown in Fig. \ref{losscompare}, the proposed loss reduces the pixel-level MSE by 6.5\% while also achieving better performance across all other evaluation metrics. This can be explained by the fact that the design loss owns the ability to explore structural similarity, which finally makes the model perform better. 
Moreover, experimental results demonstrate that the proposed CKMTransUNet model achieves an average inference time of 0.012 seconds per prediction, representing an eightfold reduction compared to the 0.098 seconds required by accelerated ray-tracing algorithms in Sionna. By significantly reducing the reliance on pilot signals for pathloss estimation, the proposed approach also enhances spectral utilization efficiency.

\begin{table}[!h]
	\centering
	\caption{Comparison of different methods in the absence of observation.}
	\setlength{\tabcolsep}{4pt}
	\renewcommand{\arraystretch}{1.5}
	\begin{tabular}{l @{\hspace{1em}} c c c c c}
		\hline
		\hline
		\multicolumn{1}{l}{Methods} & MSE & RMSE & NMSE & PSNR$\uparrow$ & SSIM$\uparrow$ \\
		\hline
		RME-GAN     & 0.003914 & 0.06257 & 0.10244 & 24.07 & 0.7713 \\
		UNet        & 0.003565 & 0.05971 & 0.09367 & 24.48 & 0.7813 \\
		RadioWNet   & 0.003161 & 0.05622 & 0.08342 & 25.00 & 0.7924 \\
		CKMTransUNet   & $\textbf{0.002105}$ & $\textbf{0.04420}$ & $\textbf{0.05546}$ & $\textbf{26.77}$ & $\textbf{0.8114}$ \\
		\hline
		\hline
	\end{tabular}
	\label{Compare_without_observation}
\end{table}

Next, we conduct ablation studies to evaluate the contributions of the CNN and transformer components within the CKMTransUNet architecture. We examine three model configurations: the full CKMTransUNet integrates 3, 4, and 9 ResNet layers in Downblocks 2, 3, and 4 with 12 transformer layers, respectively. A simplified variant, termed Easy Clayer, reduces the ResNet layers in all three Downblocks to 1 while keeping the transformer layers unchanged. While another variant, Easy Tlayer, retains the original ResNet layers but uses only 1 transformer layer. Table \ref{Ablation} summarizes the performance metrics across these configurations. The results demonstrate that both CNN and transformer layers play critical roles in the reconstruction task, with the CNN component exerting a greater influence. This can be attributed to the fact that weakened CNN layers compromise feature extraction capability, which in turn adversely affects the performance of subsequent transformer layers.

\begin{table}[!h]
	\centering
	\caption{Ablation studies to evaluate different blocks of CKMTransUNet model.} 
	\setlength{\tabcolsep}{4pt}
	\renewcommand{\arraystretch}{1.5}
	\begin{tabular}{l @{\hspace{1em}} c c c c c}
		\hline
		\hline
		\multicolumn{1}{l}{Methods} & MSE & RMSE & NMSE & PSNR$\uparrow$ & SSIM$\uparrow$ \\
		\hline
		Easy Clayer & 0.003333 & 0.05773 & 0.08801 & 24.77 & 0.7540 \\
		Easy Tlayer & 0.002337 & 0.04835 & 0.06147 & 26.31 & 0.8078 \\
		Full CKMTransUNet & $\textbf{0.002105}$ & $\textbf{0.04420}$ & $\textbf{0.05546}$ & $\textbf{26.77}$ & $\textbf{0.8114}$ \\
		\hline
		\hline
	\end{tabular}
	\label{Ablation}
\end{table}

\begin{table*}[!t]
	\centering
	\caption{Performance comparison of different methods under various observation percents for BeamCKM construction.}
	\label{performance_comparison_with_observations}
	\small
	\begin{tabular}{c*{13}{c}}
		\toprule
		\midrule
		{Percents of observation} & \multicolumn{5}{c}{0.5\%} & \multicolumn{5}{c}{1\%} \\
		\cmidrule(lr){2-6} \cmidrule(lr){7-11}
		Metrics & MSE$\downarrow$ & RMSE$\downarrow$ & NMSE$\downarrow$ & PSNR$\uparrow$ & SSIM$\uparrow$ & MSE$\downarrow$ & RMSE$\downarrow$ & NMSE$\downarrow$ & PSNR$\uparrow$ & SSIM$\uparrow$ \\
		\midrule
		Kriging & 0.007939 & 0.0891 & 0.2331 & 21.0023 & 0.4966 & 0.007189 & 0.0848 & 0.2101 & 21.4333 & 0.5151 \\
		UNet & 0.001559 & 0.0395 & 0.0391 & 28.0715 & 0.8522 & 0.001135 & 0.0337 & 0.0283 & 29.4500 & 0.8798 \\
		RadioWNet & 0.001424 & 0.0377 & 0.0354 & 28.4649 & 08626 & 0.000997 & 0.0316 & 0.0246 & 30.0130 & 0.8913 \\
		RME-GAN & 0.001355 & 0.0368 & 0.0339 & 28.6806 & 0.8624 & 0.000989 & 0.0314 & 0.0246 & 30.0480 & 0.8886 \\
		$\text{M}^3$ChanNet(1) & 0.001089 & 0.0330 & 0.0271 & 29.6297 & 0.8836 & 0.000838 & 0.0289 & 0.0207 & 30.7676 & 0.9024 \\
		$\text{M}^3$ChanNet(2) & \textbf{0.001074} & \textbf{0.0328} & \textbf{0.0267} & \textbf{29.6900} & \textbf{0.8838} & \textbf{0.000808} & \textbf{0.0284} & \textbf{0.0199} & \textbf{30.9259} & \textbf{0.9053} \\
		\midrule
		\midrule
		{Percents of observation} & \multicolumn{5}{c}{2\%} & \multicolumn{5}{c}{3\%} \\
		\cmidrule(lr){2-6} \cmidrule(lr){7-11}
		Metrics & MSE$\downarrow$ & RMSE$\downarrow$ & NMSE$\downarrow$ & PSNR$\uparrow$ & SSIM$\uparrow$ & MSE$\downarrow$ & RMSE$\downarrow$ & NMSE$\downarrow$ & PSNR$\uparrow$ & SSIM$\uparrow$ \\
		\midrule
		Kriging & 0.006664 & 0.0816 & 0.1946 & 21.7627 & 0.5285 & 0.006400 & 0.0800 & 0.1868 & 21.9382 & 0.5359 \\
		UNet & 0.000814 & 0.0285 & 0.0201 & 30.8938 & 0.9051 & 0.000670 & 0.0259 & 0.0164 & 31.7393 & 0.9179 \\
		RadioWNet & 0.000726 & 0.0269 & 0.0177 & 31.3906 & 0.9133 & 0.000615 & 0.0248 & 0.0149 & 32.1112 & 0.9234 \\
		RME-GAN & 0.000717 & 0.0268 & 0.0176 & 31.4448 & 0.9111 & 0.000600 & 0.0245 & 0.0146 & 32.2185 & 0.9216 \\
		$\text{M}^3$ChanNet(1) & 0.000670 & 0.0259 & 0.0167 & 31.7393 & 0.9189 & 0.000564 & 0.0237 & 0.0138 & 32.4872 & 0.9286 \\
		$\text{M}^3$ChanNet(2) & \textbf{0.000625} & \textbf{0.0250} & \textbf{0.0153} & \textbf{32.0412} & \textbf{0.9214} & \textbf{0.000540} & \textbf{0.0232} & \textbf{0.0131} & \textbf{32.6761} & \textbf{0.9295} \\
		\midrule
		\midrule
		{Percents of observation} & \multicolumn{5}{c}{4\%} & \multicolumn{5}{c}{5\%} \\
		\cmidrule(lr){2-6} \cmidrule(lr){7-11}
		Metrics & MSE$\downarrow$ & RMSE$\downarrow$ & NMSE$\downarrow$ & PSNR$\uparrow$ & SSIM$\uparrow$ & MSE$\downarrow$ & RMSE$\downarrow$ & NMSE$\downarrow$ & PSNR$\uparrow$ & SSIM$\uparrow$ \\
		\midrule
		Kriging & 0.006212 & 0.0788 & 0.1812 & 22.0677 & 0.5417 & 0.006063 & 0.0778 & 0.1769 & 22.1731 & 0.5469 \\
		UNet & 0.000612 & 0.0247 & 0.0150 & 32.1325 & 0.9228 & 0.000558 & 0.0236 & 0.0137 & 32.5337 & 0.9294 \\
		RadioWNet & 0.000537 & 0.0232 & 0.0130 & 32.7003 & 0.9304 & 0.000513 & 0.0226 & 0.0127 & 32.8988 & 0.9334 \\
		RME-GAN & 0.000528 & 0.0230 & 0.0128 & 32.7737 & 0.9288 & 0.000488 & 0.0221 & 0.0119 & 33.1158 & 0.9343 \\
		$\text{M}^3$ChanNet(1) & 0.000510 & 0.0226 & 0.0124 & 32.9243 & 0.9339 & 0.000470 & 0.0217 & 0.0113 & 33.2790 & 0.9380 \\
		$\text{M}^3$ChanNet(2) & \textbf{0.000489} & \textbf{0.0221} & \textbf{0.0118} & \textbf{33.1069} & \textbf{0.9341} & \textbf{0.000450} & \textbf{0.0212} & \textbf{0.0107} & \textbf{33.4679} & \textbf{0.9392} \\
		\midrule
		\bottomrule
	\end{tabular}
\end{table*}

\begin{figure}[!h]
	\centering
	\hspace*{-0.5cm}
	\includegraphics[scale=0.35]{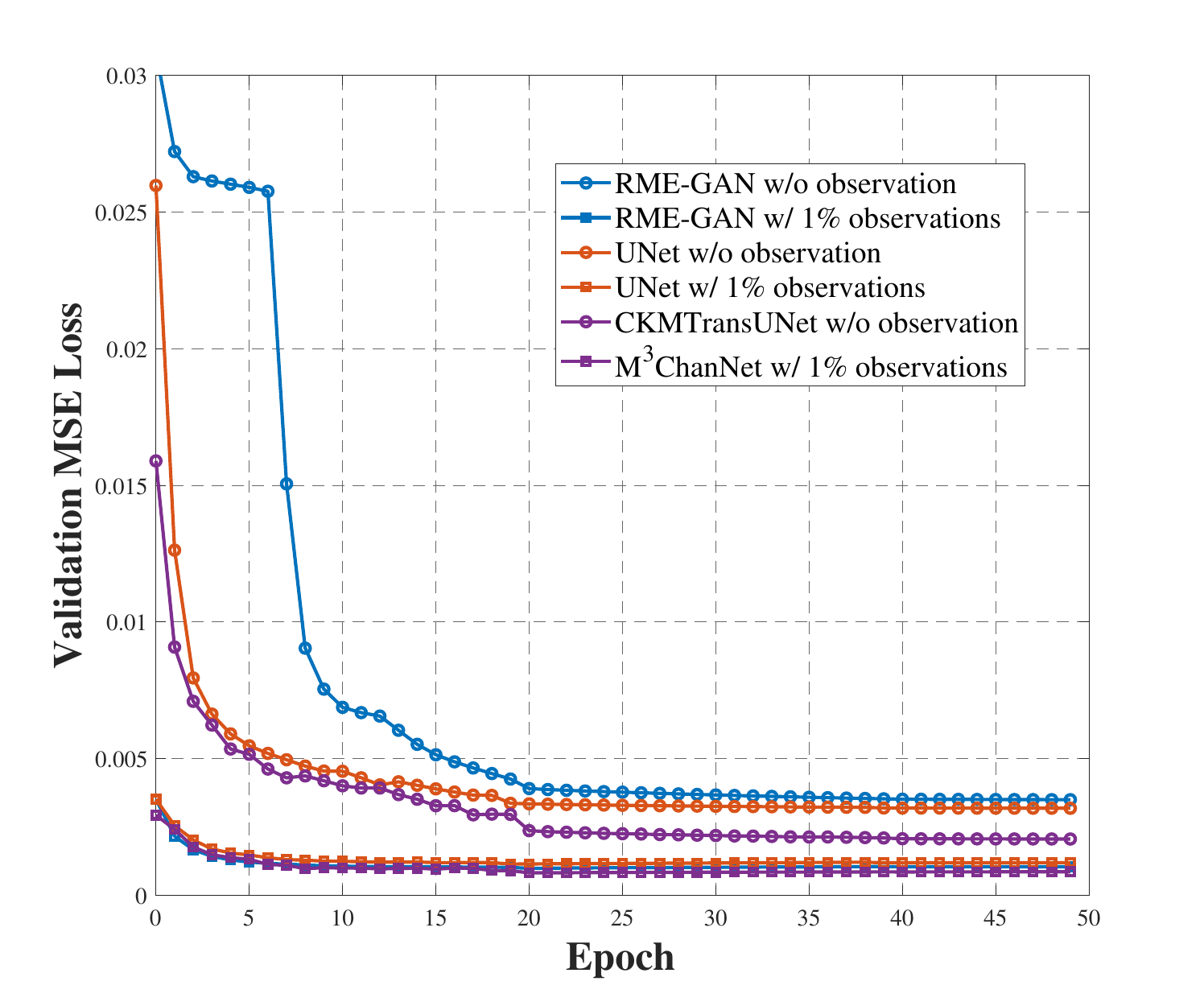} 
	\caption{Validation loss against training epoch.}
	\label{trainProcedure}
\end{figure}

\begin{figure}[!t]
	\centering
	\includegraphics[scale=0.4]{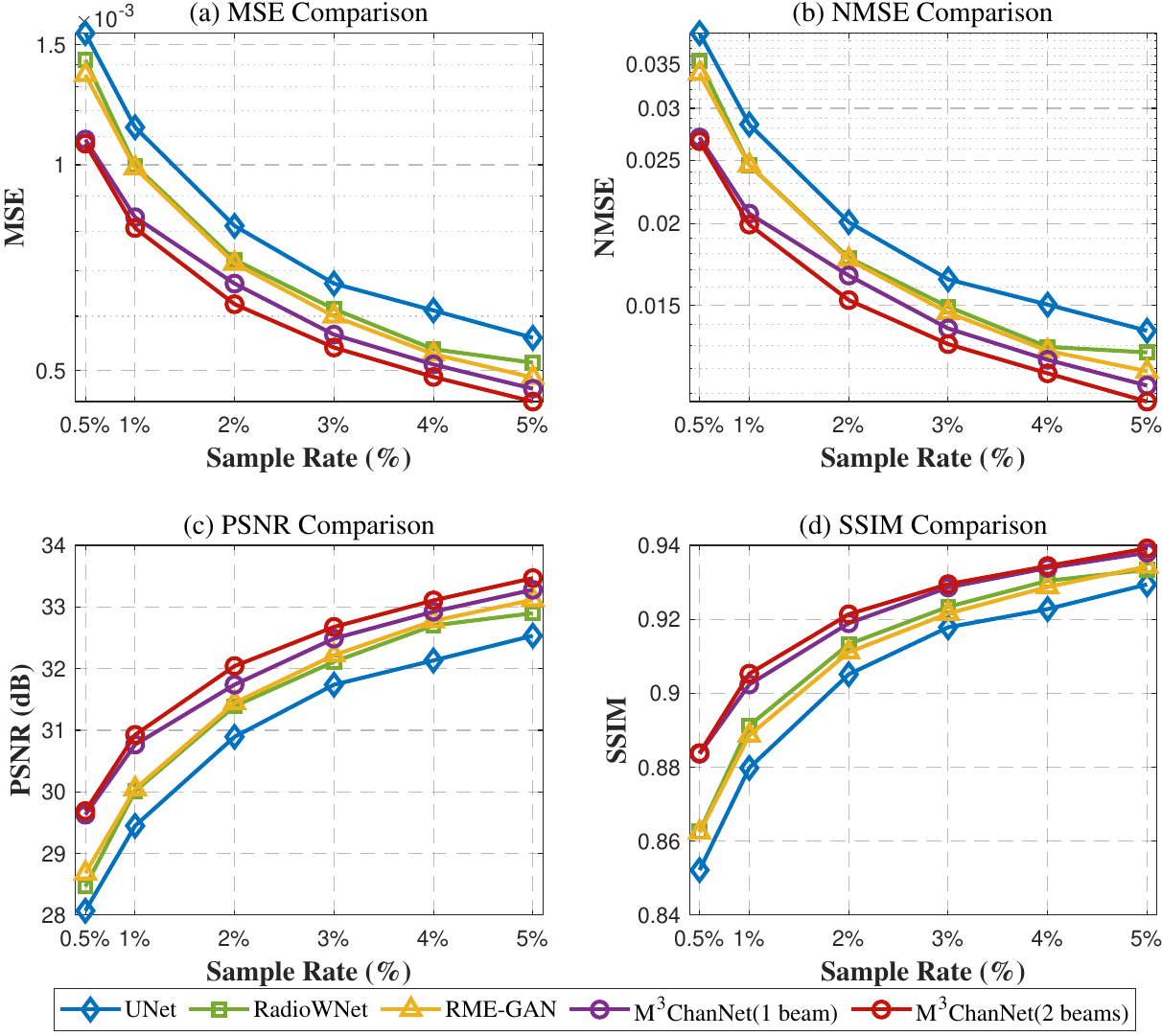} 
	\caption{BeamCKM construction performance evaluation against observation samping rates for different networks.}
	\label{combine}
\end{figure}

\subsection{Simulation with Observations}
In this section, we evaluate the $\text{M}^3$ChanNet model under sparse observations. The inputs to all compared models include observations from the target beam $\mathcal{S}_{\text{tag}}$. For the $\text{M}^3$ChanNet model, we further augment its input by randomly incorporating observations from an adjacent beam pattern to evaluate the effectiveness of the cross-attention mechanism. As shown in Fig. \ref{trainProcedure}, which illustrates the training process, the network converges more rapidly to a lower loss value when beam observations are provided. The results of different methods are presented in Fig. \ref{combine} and Table \ref{performance_comparison_with_observations}, where $\text{M}^3$ChanNet(1) means only target beam observation is provided, and $\text{M}^3$ChanNet(2) means another adjacent beam observation is provided, demonstrating that $\text{M}^3$ChanNet consistently outperforms existing network architectures. Moreover, when observations from adjacent beam are incorporated, the evaluation metrics show further improvement. The benefit conferred by the cross-attention mechanism becomes more noticeable as the number of observations increases.

Due to the inherent inaccuracies in environmental information provided by GPS, the building contour maps may not be precise, which can consequently lead to degraded model performance \cite{ckmdiff}. To further investigate the robustness of different models, we conducted tests by adding Gaussian noise to the environmental contour maps to simulate location errors. We applied zero-mean Gaussian noise with $\sigma_N^2=1$ to the binary environmental contour maps and used a clip function to constrain the value range of the input image to $0 \sim 1$. Under a 5\% sampling rate as the conditioning input, the performance of different models before and after noise injection is presented in Table \ref{robustness}. The experimental results reveal that following the introduction of Gaussian noise, the MSE values of all values are increased. It is noteworthy that $\text{M}^3$ChanNet achieves the best performance under both noise-free and noisy conditions, outperforming RME-GAN by 3.7\% in the absence of noise and by 6.2\% in the presence of noise, demonstrating superior robustness.

\begin{table}[!h]
	\centering
	\caption{Robustness evaluation for different networks.}
	\setlength{\tabcolsep}{3pt}
	\renewcommand{\arraystretch}{1.5}
	\begin{tabular}{l @{\hspace{1em}} *{5}{c}}
		\hline
		\hline
		\multicolumn{1}{l}{Metrics} & MSE & RMSE & NMSE & PSNR & SSIM \\
		\hline
		UNet w/o noise & 0.000558 & 0.0236 & 0.0137 & 32.5337 & 0.9294 \\
		UNet w/ noise & 0.000688 & 0.0262 & 0.0169 & 31.6242 & 0.9231 \\
		\hline
		RadioWNet w/o noise & 0.000513 & 0.0226 & 0.0127 & 32.8988 & 0.9334 \\
		RadioWNet w/ noise & 0.000684 & 0.0261 & 0.0167 & 31.6489 & 0.9231 \\
		\hline
		RME-GAN w/o noise & 0.000488 & 0.0211 & 0.0119 & 33.1158 & 0.9380 \\
		RME-GAN w/ noise & 0.000684 & 0.0262 & 0.0167 & 31.6500 & 0.9240 \\
		\hline
		$\text{M}^3$ChanNet w/o noise & 0.000470 & 0.0218 & 0.0113 & 33.2790 & 0.9380 \\
		$\text{M}^3$ChanNet w/ noise & \textbf{0.000642} & \textbf{0.0253} & \textbf{0.0156} & \textbf{31.9268} & \textbf{0.9300} \\
		\hline
		\hline
	\end{tabular}
	\label{robustness}
\end{table}

\begin{figure}[!h]
	\centering
	\hspace*{0cm}
	\includegraphics[scale=0.50]{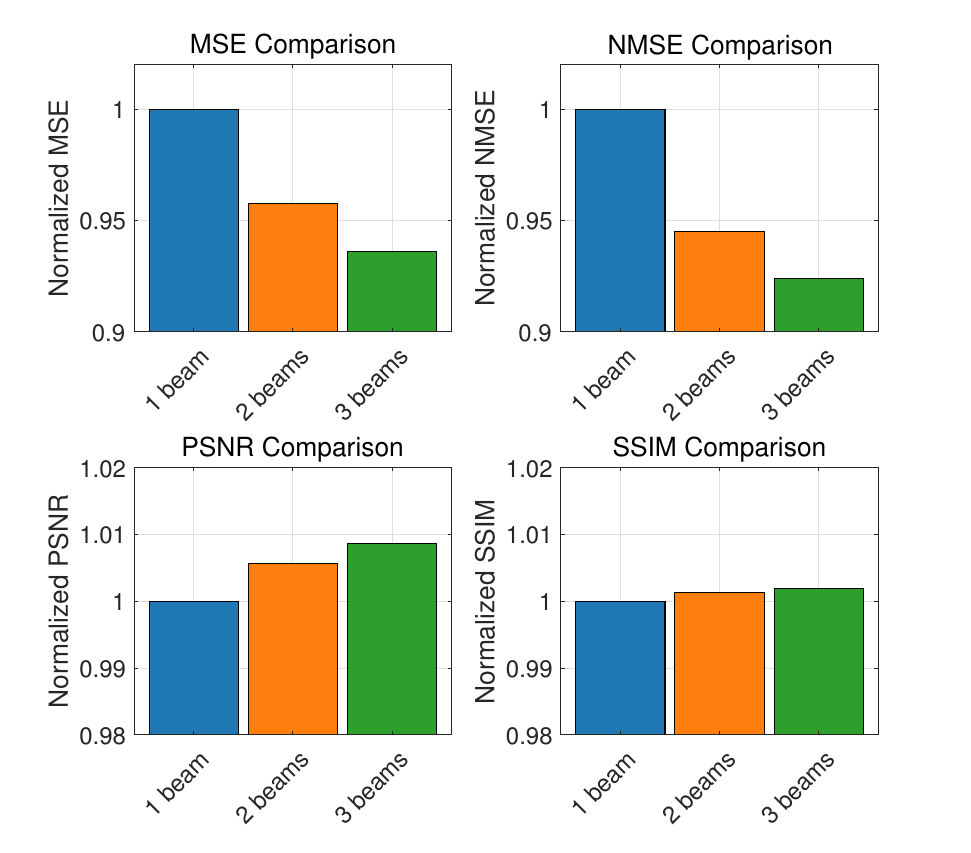} 
	\caption{Comparison chart of beam input number and metrics.}
	\label{tendency}
\end{figure}

Finally, we investigated the performance gains from incorporating additional beam patterns. As shown in Fig. \ref{tendency}, at a 5\% sampling rate, a single auxiliary beam reduced the MSE by 5\%, while a second auxiliary beam yielded a total reduction of nearly 7\%. While performance improves with more beams, the marginal benefit diminishes. This trend arises because adjacent beams provide highly relevant information, whereas more distant beams offer less similarity. Consequently, the number of auxiliary beams should be optimized to balance performance gains against informational relevance.

\section{Conclusions and Future Directions}
In this paper, we proposed the concept of BeamCKM. This model is capable of directly generating pathloss images for different DFT codewords, given the BS position and environmental contour. We also introduced a network based on the CKMTransUNet architecture. This network demonstrated superior performance over existing methods across multiple evaluation metrics. Furthermore, the study investigated the use of sparse observational data as conditional inputs to enhance BeamCKM constructions. By leveraging sparse observations from different beams, we optimized the network's performance and addressed scenarios with noisy inputs, thereby improving the model's robustness. 

Future research directions emerging from this work encompass several promising avenues for communication systems:

\noindent \textbf{BeamCKM-Assisted beamforming optimization}: The reconstructed BeamCKMs establish a spatial basis for predictive beam management. This approach enhances the efficiency of beamforming, particularly in high-mobility or non-cooperative scenarios.

\noindent \textbf{BeamCKM-enabled CSI acquisition}:
BeamCKMs provide a spatial prior to CSI acquisition and beam training. By revealing the relationship between beams and locations, they enable a focused search within a reduced candidate set, significantly lowering the overhead of exhaustive scanning and facilitating rapid link configuration in dynamic environments.

\noindent \textbf{CKM-based UAVs optimization}: UAVs provide distinct advantages for adaptive CKM construction and utilization owing to their inherent mobility and flexible deployment capabilities. Future research can investigate integrated frameworks that leverage BeamCKMs for UAV deployment, trajectory planning, and obstacle avoidance.

\balance

\bibliographystyle{IEEEtran}
\bibliography{reference.bib}

\end{document}